# An invertible crystallographic representation for general inverse design of inorganic crystals with targeted properties

Zekun Ren[1,2,*], Siyu Isaac Parker Tian[1,2,*], Juhwan Noh[3], Felipe Oviedo[4,13], Guangzong Xing[5], Jiali Li[6], Qiaohao Liang[7], Ruiming Zhu[8,9], Armin G.Aberle[2], Shijing Sun[4,14], Xiaonan Wang[6,15], Yi Liu[10], Qianxiao Li[11], Senthilnath Jayavelu[12], Kedar Hippalgaonkar[8,9], Yousung Jung[3], Tonio Buonassisi[1,4]

[1]Low Energy Electronic Systems (LEES), Singapore-MIT Alliance for Research and Technology (SMART), Singapore 138602, Singapore
[2]Solar Energy Research Institute of Singapore (SERIS), National University of Singapore, Singapore 117574, Singapore
[3]Department of Chemical and Biomolecular Engineering (BK21 Four), Korea Advanced Institute of Science and Technology (KAIST), 291 Daehakro, Daejeon 34141, Korea
[4]Department of Mechanical Engineering, Massachusetts Institute of Technology, Cambridge, MA 02139-4307, USA
[5]Okinawa Institute of Science and Technology Graduate University (OIST), Okinawa 904-0495, Japan
[6]Department of Chemical and Biomolecular Engineering, National University of Singapore, Singapore 117585, Singapore
[7]Department of Materials Science and Engineering, Massachusetts Institute of Technology, Cambridge, MA 02139, USA
[8]Institute of Materials Research and Engineering, Agency for Science, Technology and Research (A∗STAR), Singapore 138634, Singapore
[9]Department of Materials Science and Engineering, Nanyang Technological University, Singapore 117575, Singapore
[10]Materials Genome Institute (MGI), Shanghai University, Shanghai 200444, China
[11]Department of Mathematics, National University of Singapore, Singapore 117543, Singapore
[12]Institute for Infocomm Research, Agency for Science, Technology and Research (A∗STAR), Singapore 138632, Singapore

[*]These authors contributed equally
[13]Present address: Microsoft AI for Good, Redmond, WA 98052, USA
[14]Present address: Toyota Research Institute, Los Altos, CA 94022, USA
[15]Present address: Department of Chemical Engineering, Tsinghua University, Beijing 100084, China
Correspondence to: Zekun Ren <dannyzekunren@gmail.com>, Tonio Buonassisi <buonassi@mit.edu>.

## Abstract

Realizing general inverse design could greatly accelerate the discovery of new materials with user-defined properties. However, state-of-the-art generative models tend to be limited to a specific composition or crystal structure. Herein, we present a framework capable of general inverse design (not limited to a given set of elements or crystal structures), featuring a generalized invertible representation that encodes crystals in both real and reciprocal space, and a property-structured latent space from a variational autoencoder (VAE). In three design cases, the framework generates 142 new crystals with user-defined formation energies, bandgap, thermoelectric (TE) power factor, and combinations thereof. These generated crystals, absent in the training database, are validated by first-principles calculations. The success rates (number of first-principles-validated target-satisfying crystals/number of designed crystals) ranges between 7.1% and 38.9%. These results represent a significant step toward property-driven general inverse design using generative models, although practical challenges remain when coupled with experimental synthesis.

## Introduction

A common quest in materials research is to create a new material with a combination of user-specified properties, which is not present in any materials-property database. Historically, we may recruit an experienced scientist to use their intuition, to create a list of candidate compounds using heuristics (*e.g.*, elemental substitution following a given set of rules). With the advent of materials-property databases, materials screening became commonplace. Theoretical screening of solid-state materials using elemental substitution and mixing has allowed for the discovery of several crystals with user-defined functional properties, *e.g.*, perovskite materials with tailored bandgaps.[1-6] However, even under the high-performance computing (HPC) framework, the computational cost of density functional theory (DFT) calculations is high, prohibiting an exhaustive search of the theoretical materials space.[7,8] Consequently, the leading databases contain on the order of $10^5$–$10^6$ calculations for solid materials[9-11] — only a tiny fraction of the number of stoichiometric inorganic compounds believed to be possible (order $10^{10}$ considering quaternary crystals).[12,13]

To overcome these limitations, machine-learning (ML) methods have been developed to inversely design crystalline solids. (Inverse design refers to the act of a user defining target material properties and inferring a material that meets target properties, *e.g.*, by using an algorithm).[14,15] For the ML model, there are two major approaches to inversely design crystalline solids used today: global optimization and generative models. Global optimization, also called "directed evolution," involves modifying known or randomly enumerated compounds using a set of rules to design new compounds; their exploratory capacity is limited by the initial selection of structures and elements. Generative models, which learn a given data distribution, directly model all training-set materials into a probabilistic representation, from which new materials can be sampled.

Two most commonly used generative models in inverse design of solid-state materials are generative adversarial network (GAN)[16] and variational autoencoder (VAE)[17], and the key enabler of both, on top of the algorithm itself, is an invertible crystallographic representation. By "invertible", we refer to the both-way conversion from materials to representation, and vice versa. Especially, the conversion from representation to materials (the inverse of the one-way conversion from material to representation, enabling property prediction) requires a materials representation that enables the algorithm to extract necessary crystallographic information, for instance, the site location and occupancy information contained in a crystallographic information file (CIF), an input file format of choice for DFT calculations.

Because of the difficulty in creating a general invertible crystallographic representation, early demonstrations of inverse design using generative models were often limited to a fixed subset of elements[5,6,13,18,19] or crystal structures[20-22]. We observe that there lacks a general inverse design framework using generative models for inorganic crystals in prior art. Our definition of "general inverse design" is the ability to produce a prediction of a specific material (both chemistry and structure) on the basis of a user-specified target property (or properties), *i.e.*, solving the inverse problem of property prediction. To perform inverse design using ML, as opposed to human intuition, we posit that a materials representation is required that must include two key elements: (1) a representation that incorporates both structure and chemical composition into the descriptor (both structure and composition varying), and (2) a representation that is invertible, making it amenable to solving the inverse problem (property → structure + chemistry). Current invertible representations (Figure 1, and Table S1) are representative of either only specific compositional spaces (not composition varying), such as $V_xO_y$ space[5,6,13,18,19], or only specific structural spaces (not structure varying), such as cubic structures.[20-22] Korolev *et al.* proposed a spectrum representation of composition + powder X-ray diffraction (XRD) pattern;[23] although allowing variation of both composition and structure, the conversion from representation to material, *i.e.*, the construction of a unit cell from the composition + powder XRD pattern, is hard to achieve with algorithmic automation, but may still be possible for experienced human experts (thus our rating of "Limited/No" for its invertibility in Table S1). An outstanding challenge in the field is to develop an invertible crystallographic representation that accesses various chemical systems (composition varying) and various crystallographic space groups (structure varying), thus enabling general (and property-driven) inverse design.

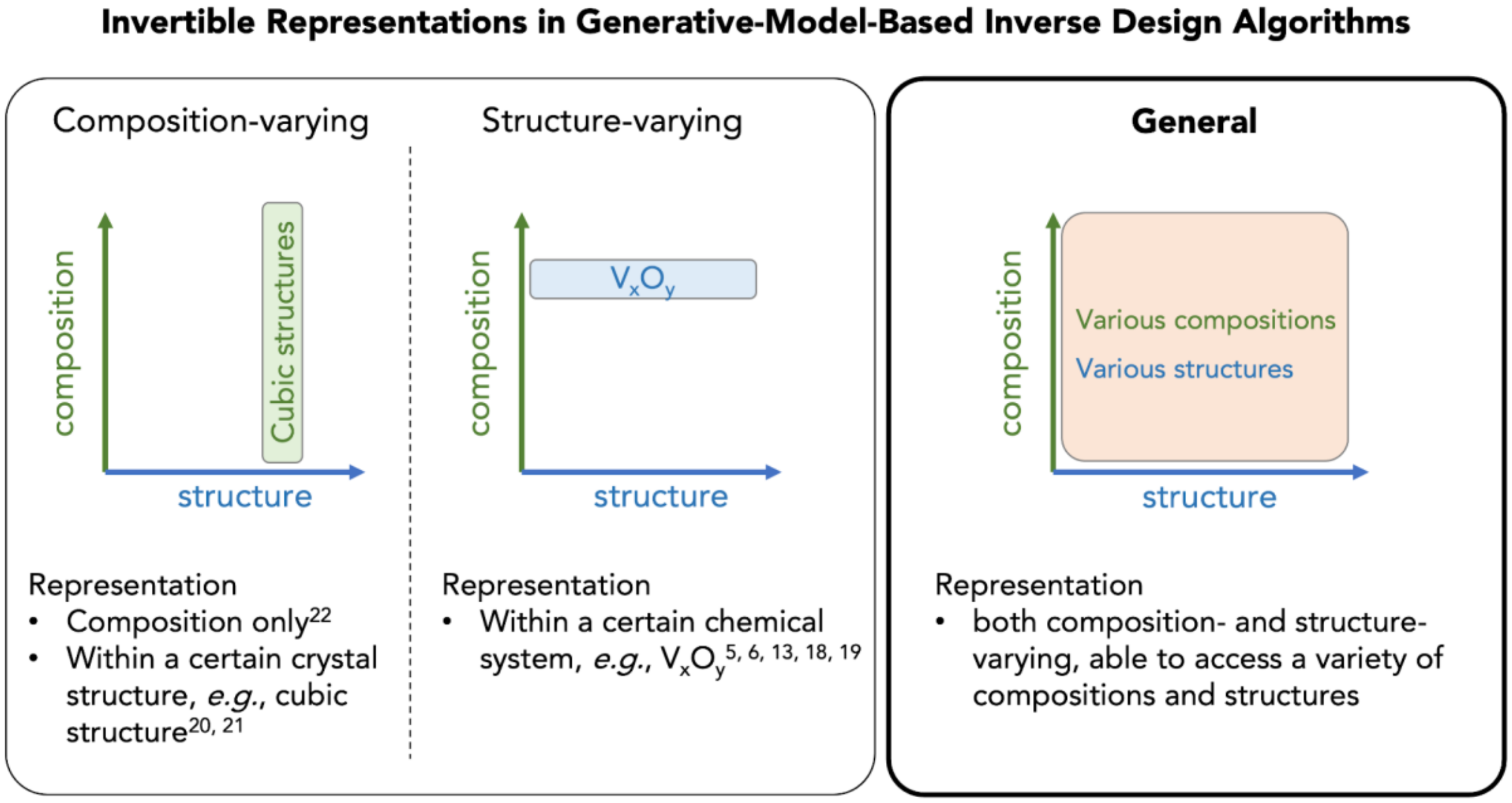

Figure 1. Novelty of the FTCP inverse design framework (right), compared to prior art (left)

The FTCP framework allows for both composition and structure to vary, enabling general and property-driven inverse design. See also Table S1.

In this work, we propose a framework for general inverse design of inorganic crystals, which is orders-of-magnitude faster than naive screening of the entire chemical space. The framework features two main components: (1) A generalized crystallographic representation, which is both composition and structure varying. The representation comprises (A) real-space features, containing CIF-like information, thus guaranteeing the invertibility, and (B) reciprocal-space features, a reciprocal-space formulation of crystal properties similar to the structure-factor (Fourier transform) calculation in XRD, as an additional featurizer. The latter is inspired by Ziletti *et al.*'s representation of 2D diffraction fingerprint,[24] which leverages the more compact crystal periodicity and symmetries in the reciprocal space. (2) A VAE, with an extra target-learning branch connected to the latent space, mapping latent vectors/points to user-specified property(ies); the encoder encodes the crystals in the training set into a continuous probabilistic latent space, and the decoder decodes any vector in the latent space to its corresponding crystals (sampling the latent space for designing new crystals); the target-learning branch during training will jointly organize the latent space to reflect a continuous change in property, *i.e.*, a property gradient, and we term thus-formed latent space "property structured." We name both the generalized invertible representation (1) and the overall framework (1)+(2) after the reciprocal-space featurizer, as the Fourier-transformed crystal properties (FTCP) framework (sometimes simply referred to as FTCP).

Using FTCP, we demonstrate three inverse design cases with design targets from single to multiple, from simple to complex: (1) case 1, designing for various targeted formation energies, $E_f$ (ranging from −0.3 to −0.7 eV/atom); (2) case 2, designing for bandgap, $E_g$ = 1.5 eV, while keeping $E_f < -1.5$ eV/atom (a desirable bandgap target for photovoltaic and optoelectronic applications); (3) case 3, designing for thermoelectric (TE) power factor (an excited-state property) to be as large as possible, while keeping $0.3 < E_g < 1.5$ eV and $E_f < 0$ eV/atom (a desirable power factor target for high-efficiency TE materials and a desirable bandgap range for a low- and medium-temperature range). The designed crystals are unique, *i.e.*, not in the Materials Project[9] database (from where our training and test sets derive), and they span a variety of chemistries and crystal structures (*i.e.*, both composition and structure varying). We "validate" the designed crystals using DFT structural relaxation followed by property prediction using first-principles calculations and define a designed crystal as "successful" if it relaxes properly and achieves the user-specified target property to within a given range. To compare FTCP against a baseline, we calculate the "random success rate" as the probability of finding a material with the user-specified target property by randomly picking from the training set, and we quantify the improvement of FTCP over random.

A summary of main results presented in the paper follows: (1) case 1, FTCP achieves success rates (number of target-satisfying designed crystals/number of total designed crystals) from 14.3% to 38.9%, scoring improvement over random from 38.8% to 270%; (2) case 2, success rate is 36.8%, achieving improvement over random of 560%; (3) case 3, FTCP designs two unique crystals that achieve comparable peak power factors with the state-of-the-art TE material, germanium telluride (GeTe), scoring a success rate of 7.1% (improvement over random not quantified due to the lack of complete power factor labels).

With the above cases, we demonstrate the usefulness of FTCP and validate using first-principles calculations, as is common practice in prior art (references are detailed in the section synthesizability challenge). We posit that a complete experimental validation is outside of scope for this study, but to lay the groundwork toward the experimental-synthesis goal of inverse design, we explore adding a näive synthesizability metric, *i.e.*, the existence or absence of an Inorganic Crystal Structure Database (ICSD) entry (previously explored by Jang *et al.*),[25,26] to the target-learning branch. This organizes the property-structured latent space according to a synthesizability metric, which can direct generative-design sampling toward more favorable regions (of higher likelihood to possess an ICSD entry). We conclude with a brief discussion of the invariance challenge faced by structure-conscious invertible crystallographic representations, including the FTCP representation.

# Results and Discussion

## Inverse Design Framework, FTCP

### Representation

Representing an infinitely repeating 3D crystal is hard, compared with representing small organic molecules, due to the periodicity, the complex geometrical symmetries (230 space groups for 3D periodic crystals), and the vast chemical space (up to 81 nonradioactive elements). To sufficiently capture this variety of crystals while satisfying invertibility, we propose a crystallographic representation of real-space CIF-like features combined with reciprocal-space Fourier-transformed features (Figure 2A).

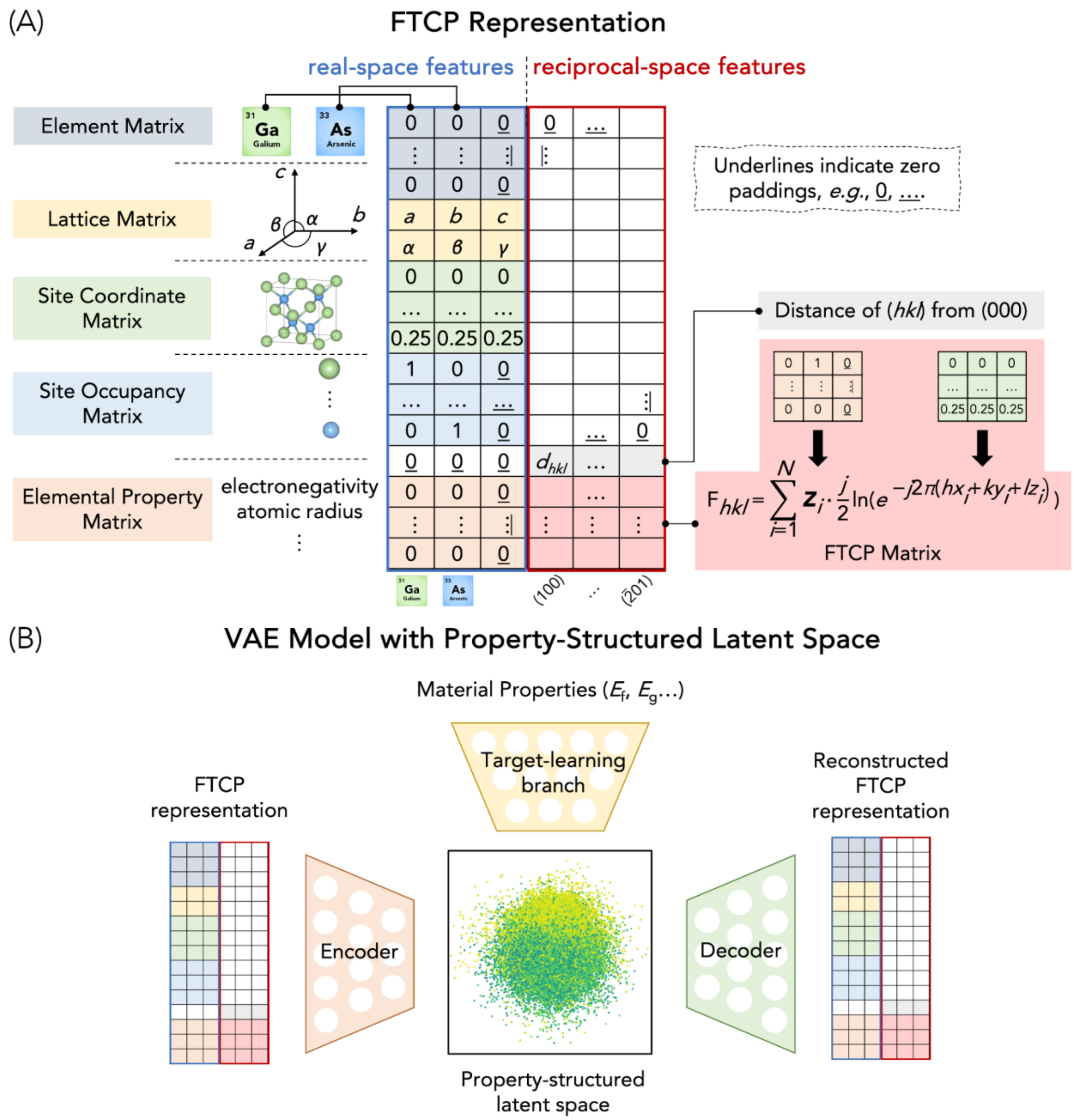

Figure 2. FTCP framework, representation + model

(A) Representation: The invertible FTCP representation contains both real- and reciprocal-space features. Real-space contain CIF-like features, such as the element matrix (describing constituent elements), the lattice matrix (describing lattice parameters), the site coordinate matrix (describing fractional coordinates of sites), the site occupancy matrix (describing elemental occupancy per site), as well as the elemental property matrix (elemental descriptors). Reciprocal-space features project the elemental descriptor $Z_i$ ($i$ for each site) for all the $N$ sites in the unit cell along various spatial frequencies, *hkl* (Miller indices) via spatial discrete Fourier transform (with a data preprocessing of $\frac{j}{2}\ln$ where $j$ is the imaginary unit) to form the FTCP matrix. The distance of each *hkl* ($k$ point) from (000) (light-gray row, with "$d_{hkl}$", in reciprocal-space features) is also recorded and prepended to the FTCP matrix (red box in reciprocal-space features). (GaAs unit cell is used as an example for illustration).

(B) Model: the VAE architecture using the invertible FTCP representation for inverse design. On top of the encoder + decoder architecture of a normal VAE, the latent space is also connected to a target-learning branch for property mapping, reflecting a property gradient(s) (property-structured latent space). The latent space is visualized by plotting two dimensions (the third and the 12$^{th}$) out of a total of 256. A detailed discussion of the visualization can be found in section S2.1 in supplemental information.

The real-space features contain necessary information for unit cell construction in 2D matrix form, namely a vertical concatenation of:

- Element matrix, where each column is a one-hot vector representing a constituent element.
- Lattice matrix, where the lattice parameters, $a$, $b$, $c$, $\alpha$, $\beta$, and $\gamma$, form a 2 × 3 matrix.
- Site coordinate matrix, where each row vector contains the fractional coordinate of a site (vacant sites have all-zero entries, namely zero padding).
- Site occupancy matrix, or site index matrix, where each row is a one-hot vector for a site indicating which constituent element occupies the site (vacant sites have all-zero entries).
- Elemental property matrix, where each column is an elemental property vector/elemental descriptor $\boldsymbol{Z}$ (equivalent to the atom feature vector used in the crystal graph convolutional neural network (CGCNN)[27] representation) of the constituent element, including group number, period number, electronegativity, covalent radius, and valence electrons, among others.

The real-space features guarantee invertibility because if the real-space features are set, *i.e.*, filled with values, a crystal can always be constructed. The resultant crystal may not valid, but there is always a corresponding resultant crystal. Analogous to a CIF, the values in the real-space representation always correspond to a specific crystal, although the crystal itself may not be valid.

The reciprocal-space (also called *k*-, momentum-, frequency-, and Fourier-space) features contain the spatial Fourier-transformed elemental property vectors (the transformed property is termed crystal property). The formulation of the reciprocal-space features roots in the domain knowledge of solid-state physics. Atoms in a crystal are arranged in a periodic pattern. Bloch's theorem establishes that the wave function can be expressed as the product of a plane wave and a function that has the same periodicity as the crystal.[28] The spatial periodicity of the atomic arrangement, analogous to the temporal periodicity of a signal, can be analyzed in the frequency domain with Fourier transform, or in materials science terms, the reciprocal space. We build upon the discrete spatial Fourier transform used in the structure factor calculation in scattering physics and XRD to reach Equation 1:

1. Instead of transforming the atomic scatter factor to the reciprocal space, we substitute with the elemental property vector (projecting elemental properties to obtain crystal properties).
2. Apply a preprocessing ($\frac{j}{2}\ln$ increases the prediction accuracy for the formulations attempted in this study; see details in section S2.2 in supplemental information).

$$\mathbf{F}_{hkl} = \sum_{i=1}^{N} \boldsymbol{Z_i} \cdot \frac{j}{2} \ln \left(e^{-j2\pi(hx_i + ky_i + lz_i)}\right) = \sum_{i=1}^{N} \boldsymbol{Z_i} \cdot \pi(hx_i + ky_i + lz_i) \qquad \text{(Equation 1)}$$

where *hkl* are the Miller indices/spatial frequencies, $\boldsymbol{Z}_i$ is the elemental property vector for site *i* (total *N* sites in the unit cell), $x_i y_i z_i$ are the fractional coordinates of site *i*, and *j* is the imaginary unit. The FTCP, $\mathbf{F}_{hkl}$, is calculated for 59 different combinations of *hkl* (*k* points), which, as column vectors, concatenate horizontally to form an FTCP matrix. Upon prepending the distance of the *k* points from (0, 0, 0) to the FTCP matrix, the reciprocal-space features (matrix) become complete. Horizontal concatenation of the real-space features (matrix) and the reciprocal-space ones (with zero padding) forms the FTCP crystallographic representation. (The reciprocal-space features act as an additional featurizer, which improves the reconstruction accuracy of the VAE + target-learning branch model. See details in section S2.3 in supplemental information.)

## Model

After obtaining the invertible FTCP representation (2D matrix), we construct a continuous property-structured latent space, which allows for sampling of new crystals using a VAE. The VAE comprises an encoder, a decoder, and a target-learning branch, as shown in Figure 2B. The encoder compresses the FTCP representation into a point probabilistically in the latent space of reduced dimension (with total dimensions = 256), namely the encoder outputs $\boldsymbol{z}_{\mathrm{mean}}$ and $\boldsymbol{z}_{\mathrm{variance}}$. The decoder learns to sample with $\boldsymbol{z}_{\mathrm{variance}}$ around the latent point $\boldsymbol{z}_{\mathrm{mean}}$ and to decode this vicinity of $\boldsymbol{z}_{\mathrm{mean}}$ back to the original FTCP representation. Naturally, the training for better reconstruction (minimizing reconstruction loss) results in nearness of latent points of similar FTCP representations, namely a cluster-forming behavior in the latent space for similar inputs. We also implement a standard VAE Kullback-Leibler (KL) loss to promote dense-packing of latent points around the center of the latent space, especially for those of very different FTCP representations, by encouraging $\boldsymbol{z}_{\mathrm{mean}}$ and $\boldsymbol{z}_{\mathrm{variance}}$ to follow that of a unit Gaussian. The reconstruction loss and the KL loss forms a stand VAE, where the input is embedded to a continuous probabilistic latent space — the cluster forming of the probabilistic points results in continuous change in the decoded FTCP representation, and the dense packing ensures the latent space does not contain void areas without encoded latent points. Overall, the two losses ensure a continuous latent space, which allows for easy sampling between known crystals (encoded latent points of training data). In addition, we add a feedforward target-learning branch to map latent points to certain properties, namely there is an additional property-mapping loss. (The target-learning branch handles multiple target properties by outputting a vector, instead of a scalar, containing entries for each property. Thus, the property-mapping loss is the overall loss across different properties.) This additional loss further encourages latent points to be near when they possess similar property values, thus introducing a property gradient (change) in the latent space (organizing latent points further according to the property). We term the obtained latent space "property structured." The overall loss of the VAE is a weighted sum of the three losses, *i.e.*, the reconstruction loss, the KL loss, and the property-mapping loss. The detailed architecture and the hyperparameters of the model are recorded in the section generative model and section S2.4.

## Design from trained model (sampling and postprocessing)

To sample from the trained property-structured latent space, we adopt a local perturbation (Lp) strategy, similar to that used by the decoder during training. We identify the latent points of target-satisfying crystals in the training set, and we sample around the latent points with a scaled (0.3–3) unit Gaussian noise, *i.e.*, a Gaussian noise with scaled variance. The scale controls a tradeoff between exploitation and exploration: with too small a scale, the sampled points fall too close to known crystals, displaying high reconstruction accuracy with low novelty; with too large a scale, the sampled points are more scattered, displaying high novelty with low reconstruction accuracy. Thus, our choice of the scale is determined through trial and error to maximize exploration, while ensuring sufficient exploitation by taking the largest value that yields structure design errors <20%, discussed in greater details in section S2.5 in supplemental information. Apart from Lp, we also explore two other sampling strategies, spherical linear interpolation[29] (Slerp) and global perturbation (Gp), in section S2.5 in supplemental information. We deem Lp to be the most suitable and high-achieving strategy in the setting of property-driven sampling, and we consistently use Lp in the design (use) cases in this study.

The decoding from sampled latent points results in a decoded/reconstructed FTCP representation (real-space + reciprocal-space features). The reconstruction of a unit cell lies in postprocessing the decoded real-space features, which entails the processing of the element matrix (one-hot encoded), the lattice matrix, the site coordinate matrix, and the site occupancy matrix (one-hot encoded). For one-hot encoded matrices, the respective one-hot vector is obtained by setting the maximum value in the vector to one and the rest to zero. For the site occupancy matrix, to distinguish empty sites due to zero padding, we enforce a threshold (0.05) such that empty sites are where all decoded values in the vector are below the threshold. Thus, empty sites will have all-zero vectors in the site occupancy matrix, while valid sites will have one-hot vectors after postprocessing. We then use the site occupancy matrix to ignore the decoded coordinates for empty sites.

## Workflow

Our workflow for the design cases (applying FTCP) in this study is shown in Figure 3, where we have four stages: (1) We define the target property of the intended material, (2) we design from trained model (performing sampling and postprocessing) and obtain a number of FTCP-designed candidates, (3) we pass these candidates through structural relaxation (performed by DFT in our study; in the future, this may be performed by an ML surrogate model to approximate the speed of FTCP compound generation), and we remove the nonconverged, repeated, and invalid structures, *e.g.*, with overlapping atoms. (DFT structural relaxation is a way to reduce errors in the construction of designed crystals. We discuss the sources of error and the need for structural relaxation for designed crystals in section S2.6 in supplemental information.) (4) We perform first-principles calculation(s) to verify the property(ies) of the designed candidates and retain those that satisfy the design target within a user-specified margin of error (tolerance).

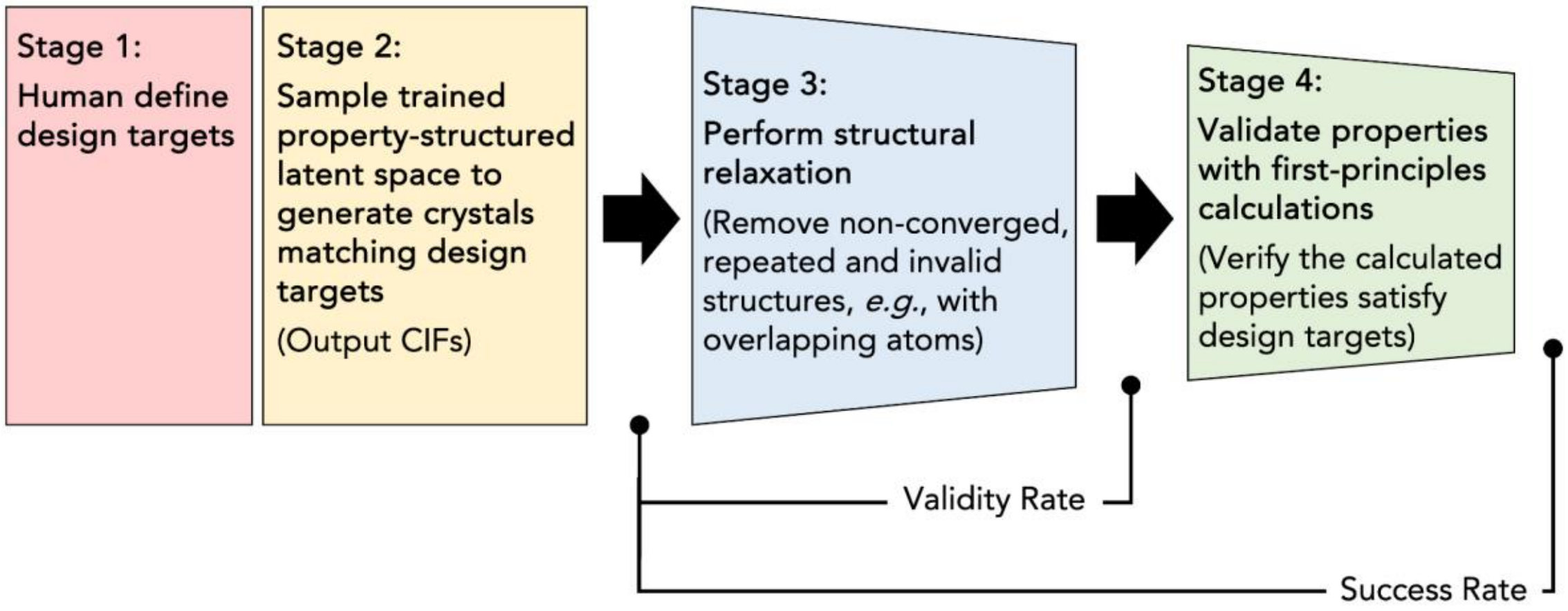

Figure 3. Workflow of the FTCP framework in design cases

Note that structural relaxation can be performed using DFT (our study), or in the future, an ML surrogate model. We define the following metrics: validity rate, percentage of FTCP-designed candidates successfully passing through structural relaxation ([no. of candidates exiting stage 3]/[no. of candidates exiting stage 2]), and success rate, percentage of FTCP-designed candidates confirmed to have the user-specified target property(ies) to within the specified margin of error ([no. of candidates exiting stage 4]/[no. of candidates exiting stage 2]).

We then define three metrics to quantify the performance of FTCP: (1) Validity rate, defined to be the percentage of FTCP-designed candidates successfully passing through the DFT structural relaxation ([no. of candidates exiting stage 3]/[no. of candidates exiting stage 2]). (2) Success rate, defined to be the percentage of FTCP-designed candidates confirmed to have the user-specified target property within tolerance ([no. of candidates exiting stage 4]/[no. of candidates exiting stage 2]). (3) Improvement over random, where random success rate is the probability of finding a material with the user-specified target property by randomly picking it from the training + test datasets [(FTCP success rate – random success rate)/random success rate]. We approximate the random success rate with the percentage of target-satisfying crystals in the training + test datasets.

With the above workflow, we apply FTCP in three design cases:

1. Design for formation energy, $E_f$: We design based on ternary crystals for four $E_f$ targets, −0.5, −0.3, −0.6, and −0.7 eV/atom, as these $E_f$ are the most prevalent in the database we use for training (Materials Project[9]). To calculate the "success metric," we define a tolerance of ±0.06 eV/atom.
2. Design for bandgap, $E_g$, with $E_f$ constraint: We design based on ternary and quaternary crystals for a design target $E_g$ = 1.5 eV, which is of interest both for solar cells (near the maximum of the detailed balance efficiency limit for single-junction devices[30] ), and for optoelectronic devices (*e.g.*, LEDs). We also set a negative $E_f$ (<−1.5 eV/atom) target as a proxy for the stability of designed crystals. To calculate the "success metric," we define a tolerance of ±0.3 eV for $E_g$ and ±0.06 eV/atom for $E_f$.
3. Design for TE power factor, with $E_g$ and $E_f$ constraints: We design based on ternary and quaternary crystals for a power factor (heat-to-electricity conversion efficiency) to be as large as possible. We also set a bandgap target between 0.3 and 1.5 eV, desirable for low- and medium-temperature range, as well as a negative (<0 eV/atom) formation energy, again as a proxy for stability. To calculate the "success metric," we compare the first-principles calculated power factor to the state of the art TE material, GeTe, and define a tolerance of ±0.3 eV for $E_g$ and ±0.06 eV/atom for $E_f$.

After obtaining the designed candidate CIFs, according to our workflow, we pass the designed candidates through DFT structural relaxation to remove nonconverged, repeated, and invalid structures (validity rate is calculated). We then apply DFT calculations for $E_f$, and $E_g$, and Boltzmann transport properties (BoltzTraP[31] ) calculations for power factor. (The details of the structural relaxation, and the first-principles calculation of properties, are described in sections S3.2 and S3.3, and we provide a list of validated FTCP-designed crystals in S3.1.) We verify the designed candidates to retain those with target-satisfying properties within aforementioned tolerances (success rate is calculated).

We summarize the performance of the design cases in Table 1 using the metrics defined above (detailed descriptions of each design case are in the following section applying FTCP: three design cases validated by first-principles calculations).

Table 1. Performance of three design cases using FTCP

| | Case 1 | | | | Case 2 | Case 3 |
|---|---|---|---|---|---|---|
| Description | Formation energy (±0.06 eV/atom tolerance) | | | | Bandgap (±0.3 eV tolerance) with formation energy constraint | Thermoelectric power factor (as high as possible) with bandgap and formation energy constraints |
| Metrics | $E_f = -0.5$ eV/atom | $E_f = -0.3$ eV/atom | $E_f = -0.6$ eV/atom | $E_f = -0.7$ eV/atom | $E_g = 1.5$ eV<br>$E_f < -1.5$ eV/atom | Power factor as large as possible<br>0.3 eV ≤ $E_g$ ≤ 1.5 eV<br>$E_f < 0$ eV/atom |
| Validity Rate | 77.8% (14/18) | 81.0% (17/21) | 96.4% (27/28) | 92.9% (26/28) | 84.2% (16/19) | 42.9% (12/28) |
| Success Rate (FTCP) | 38.9% (7/18) | 14.3% (3/21) | 17.9% (5/28) | 21.4% (6/28) | 36.8% (7/19) | 7.1% (2/28) |
| Success Rate (Random)*† | 10.5% (2781/26402) | 10.3% (2732/26402) | 9.6% (2522/26402) | 8.3% (2183/26402) | 5.5% (3035/54925) | –‡ |
| Improvement over Random† | 270% | 38.8% | 86% | 150% | 560% | –‡ |

*approximated by the percentage of target-satisfying crystals in the dataset (training + test).

†calculated with an updated version of Materials Project accessed on 14 Sep 2021. (Materials Project has updated since our design cases, which accessed Materials Project on 22 Jun 2020.)

‡not calculated for case 3 because of the lack of calculated power factor values for every crystal and the use of a qualitative criterion, "power factor as large as possible".

We observe a rule of thumb that with more design targets (increased design difficulty), the percentage of target-satisfying crystals in the dataset (random success rate) decreases, and the validity rate and the success rate also both decrease because there are more losses (property-mapping losses) to optimize. As the inverse design problem becomes more complex (*e.g.*, the number of target properties increases, the tolerance tightens), the usefulness of FTCP is expected to grow; see the improvement over random for case 2 versus case 1, for example.

There are two sources of error reducing success rate: one is from imperfect positioning of latent points and imperfect reconstruction (KL loss and reconstruction loss), and one is from imperfect property mapping (property-mapping loss), which can yield three scenarios of error: (1) The validity of the decoded crystal is too low that the geometry is wrong or the structural relaxation fails. (2) The decoded crystal successfully relaxes, but the resultant crystal is not what is intended after relaxation (thus, the first-principles calculation of the resultant crystal yields the incorrect property value). (3) The property mapping is not accurate, resulting in an inaccurate property gradient (thus, the first-principles calculation yields an incorrect property value). Scenarios 2 and 3 can also happen at the same time. We posit that although a high success rate is ideal, the main achievement of FTCP is realizing a nonzero success rate of general inverse design. Many research problems only require one or a few successful candidate(s); the minimum success rate is therefore on the order of the inverse of the throughput of the synthesis tool.

## Applying FTCP: Three design cases validated by first-principles calculations

### Case 1: Design for formation energy and case 2: Design for bandgap (with formation-energy constraint): *Case studies for photovoltaic and optoelectronic applications*

In the first design case (case 1) of the FTCP framework, we design new crystals based on ternary crystals with target $E_f = -0.5$ eV/atom. As our dataset (training + test), we select the ternary crystals of ≤20 sites with energy above hull <0.08 eV/atom in the Materials Project database. Examples of six FTCP-designed 3D crystals with targeted $E_f$ after DFT relaxation are shown in Figure 4. These crystals cannot be found in the Materials Project database, indicating their uniqueness. We perform DFT calculations of structural relaxation and formation energy using GGA(+U). The DFT-calculated formation energies are shown in Figure 5A. Seven out of 18 (38.9%) have $E_f$ meeting the target −0.5 eV/atom (within a user-specified tolerance of ±0.06 eV/atom). There are in total eight different crystal structures with more than 30 different elements in those 18 crystals (detailed in section S4.1 in supplemental information), demonstrating that the FTCP framework can access a wide range of structures and chemistries.

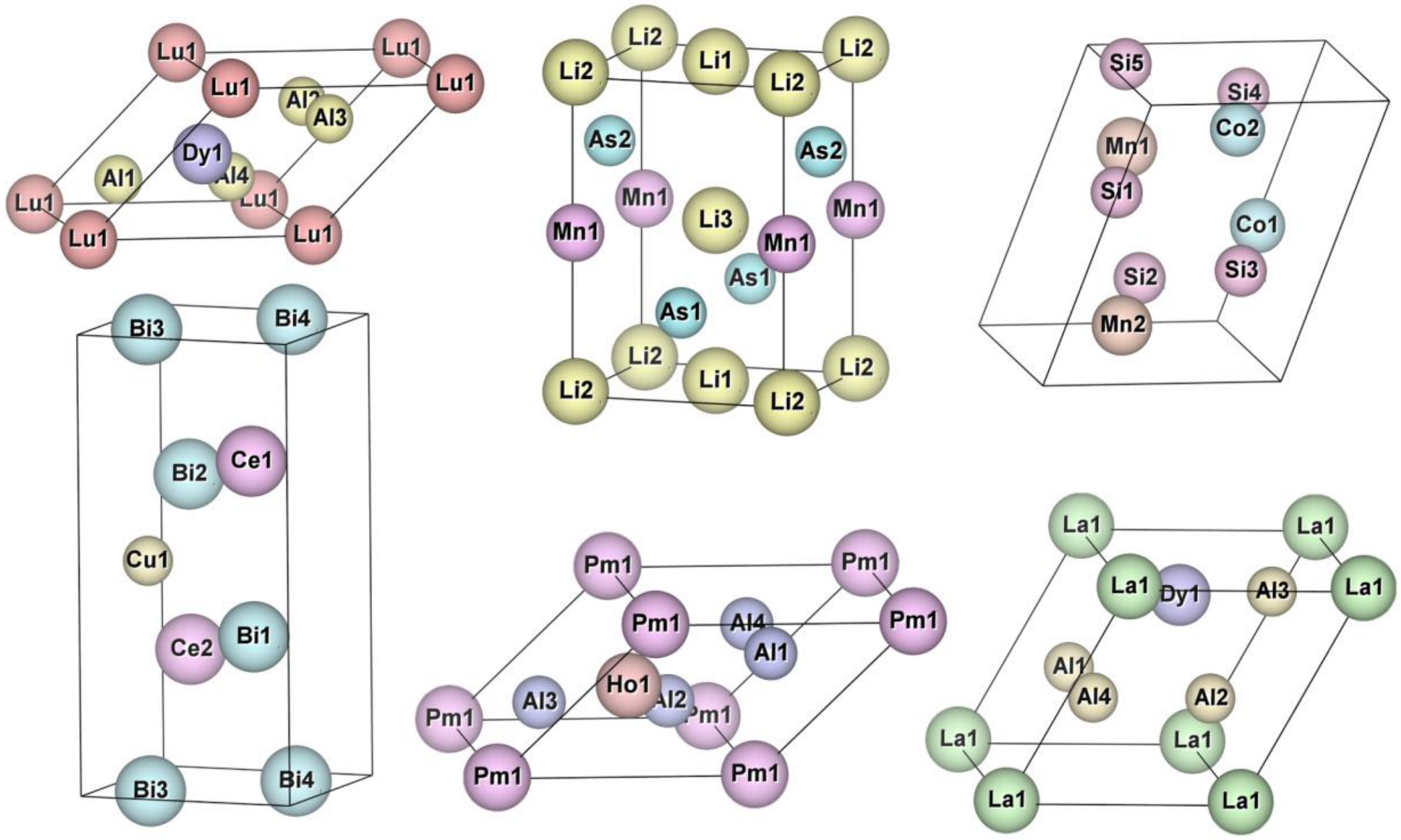

Figure 4. Examples of FTCP-designed crystals

The designed crystals are with targeted $E_f$ = −0.5 eV/atom after DFT relaxation. See also Figure S4.

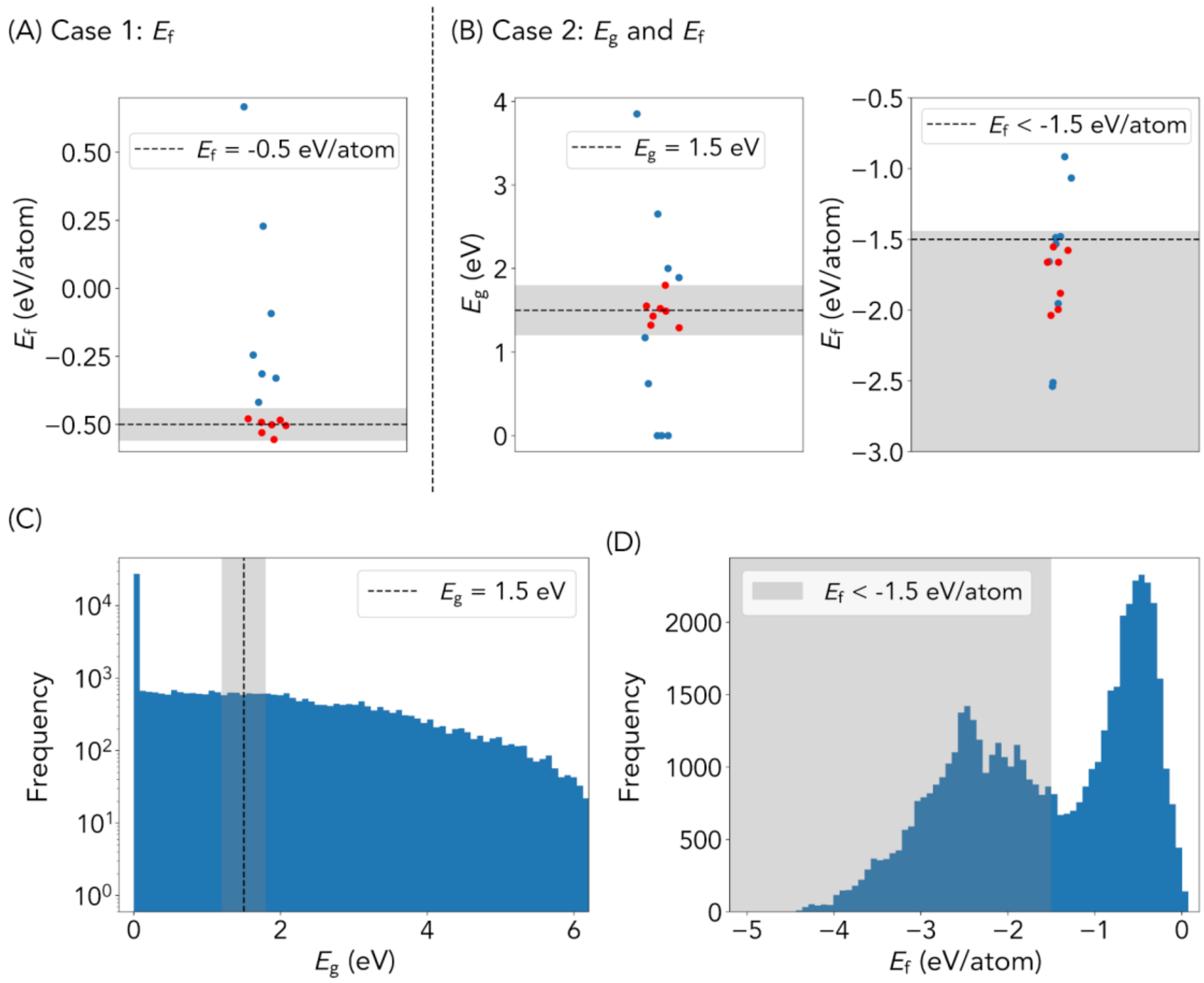

Figure 5. DFT-calculated properties of FTCP-designed crystals and target-satisfying regions in the training set

(A) Case 1: DFT-calculated $E_f$ for designed crystals based on ternary crystals with a target $E_f = -0.5$ eV/atom. The black dashed line is $E_f = -0.5$ eV/atom. Red dots indicate target-satisfying designed crystals. Seven out of 18 (38.9%) inversely designed crystals meet the target $E_f$ within tolerance (gray band of $-0.5 \pm 0.06$ eV/atom).

(B) Case 2: DFT-calculated $E_g$, and $E_f$ for designed crystals based on ternary and quaternary crystals with target $E_g$ = 1.5 eV and $E_f < -1.5$ eV/atom. The gray band indicates the target-satisfying region within tolerance. Red dots indicate target-satisfying designed crystals. Fourteen out of 19 (73.7%) of the inverse-designed crystals have $E_f < -1.5 + 0.06$ eV/atom, and seven (included in the 14) (36.8%) have $E_g = 1.5 \pm 0.3$ eV (0.06 eV/atom and 0.3 eV are user-specified tolerances for $E_f$ and $E_g$).

(C and D) Histograms of $E_g$ and $E_f$ of the crystals in the dataset of case 2, where target-satisfying crystals only comprise 5.5% of the total dataset (recorded in Table 1), indicating the low likelihood of performing inverse design by pure random chance. The gray bands indicate target-satisfying regions within tolerance. See also Figure S5.

To further test the robustness of the framework, we further inversely design another 77 crystals based on ternary crystals with staggered $E_f$ values (−0.3, −0.6, and −0.7 eV/atom). Fourteen out of those 77 solid materials have $E_f$ within the specified tolerance of ±0.06 eV/atom (section S4.2 in supplemental information). Combined with the 18 designed crystals for $E_f = -0.5$ eV/atom, we boost and evaluate the "uniqueness" of these inverse-designed crystal structures in two ways: (1) Composition: we exclude designed crystals whose compositions exist in the Materials Project database. Consequently, all the designed 84 (valid, passed DFT structural relaxation) crystals in this demonstration have unique chemical formulae that do not exist in the Materials Project database. (2) Structure: we investigate if the inverse-designed crystals can be discovered by manual elemental substitution (zero to minimum structural change). We use a structural dissimilarity value, which is calculated based on local coordination information of all sites in the two structures,[32] to assess the structural uniqueness between designed crystals and crystals in the dataset (training + test). The dissimilarity value ranges from zero to above, with zero indicating identical crystal structures, and large values (>1) indicating large dissimilarity. (The dissimilarity value does not consider elements, *e.g.*, comparing NaCl and CsF, which are both $Fm\overline{3}m$, would yield a dissimilarity value of zero.) A low dissimilarity value does not indicate similar crystals per se (such as NaCl and CsF), just similar structures, and thus, we use the dissimilarity value to assess the structural variety of the designed crystals and to suggest the likelihood of discovery by elemental substitution. For our set of 84 valid designed crystals, the median dissimilarity value is 0.37, and 11 crystals have a dissimilarity value above 0.75 (a dissimilarity threshold used in the Materials Project database). The dissimilarity value per designed crystal is taken to be the minimum dissimilarity values between the designed crystal and every crystal in the dataset (training + test), as shown in section S4.3. We posit that designed crystals with high dissimilarity values have a low likelihood of being discovered by conventional elemental substitution.

In the second design case (case 2), we extend the design criteria to multiple objectives by basing the design on ternary and quaternary crystals for bandgap $E_g$. The design criteria are: (1) bandgap = 1.5 eV (of a user-specified tolerance of 0.3 eV) and (2) formation energy < −1.5 eV/atom (of a user-specified tolerance of 0.06 eV/atom). The bandgap target is selected because it is of interest both for solar cells (near the maximum of the detailed balance efficiency limit for single-junction devices)[30] and for optoelectronic devices (*e.g.*, LEDs). The negative formation energy criterion is chosen as a crude proxy for stability of the designed crystal (see the section synthesizability challenge for a nuanced discussion). We select ternary and quaternary crystals of ≤40 sites having energy above hull <0.08 eV/atom in Materials Project as our dataset (training + test). A total of 19 crystals with unique chemical formulae are inversely designed after filtering out compositions that already exist in the Materials Project database (compositional uniqueness). We quantify the median dissimilarity value of designed crystals to be 0.57, and three out of 16 valid crystals have a dissimilarity value above 0.75 (structural uniqueness). The dissimilarity values are shown fully in section S4.3. We perform DFT validation to examine whether these designed crystals' properties meet the user-specified target (after structural relaxation). Figure 5B shows the distribution of $E_g$ and $E_f$ of the 19 designed materials. Seven out of 19 (36.8%) of the designed crystals satisfy the bandgap target $E_g$ = 1.5 (±0.3) eV, and 14, including the seven, (73.7%) satisfy the formation energy target $E_f$ < −1.5 (+0.06) eV/atom. We set the user-specified tolerance of $E_g$ of the designed crystals to be significantly larger compared to the one of $E_f$. This is due to the relatively larger prediction error for $E_g$ compared with $E_f$ in the property mapping of the target-learning branch (Table S3), on which the user-specified tolerances are based. This is attributed to the many zero values of $E_g$, a situation confusing the property mapping, and leaving the prediction for zero and near-zero values inaccurate. Figures 5C and 5D show the histogram of $E_g$ and $E_f$ values of all crystals in the dataset. The probability of finding crystals that satisfy both $E_g$ = 1.5 (±0.3) eV and $E_f$ < −1.5 (+0.06) eV/atom by random sampling (random success rate) is 5.5%, while FTCP reports a success rate of 36.8% (7/19), *i.e.*, a 560% improvement over random, establishing the nontriviality of our inverse design framework.

### Case 3: Design for TE power factor (with formation energy and bandgap constraints): *Case study for TE applications*

In the third design case (case 3), we use the FTCP framework to address one of the outstanding challenges in the field of TEs, *i.e.*, to design new earth-abundant materials that can convert heat to electrical energy and vice versa with high efficiency.[33,34] This design case is challenging from a scientific point of view, because it includes excited-state (as opposed to ground-state) properties, and from an ML point of view, because it relies on sparsely labeled training data (only a small portion of the dataset have TE-relevant property calculated). Example TE-relevant labels include carrier effective mass, Seebeck coefficient, and power factor, which are computationally expensive for DFT or require different computation platforms and are not all present in the same database.[31]

Considering the above, we set the design targets (for designing based on ternary and quaternary crystals) to be: (1) power factor as high as possible (heat-to-electricity conversion efficiency), (2) bandgap between 0.3 and 1.5 eV (desirable for low- and medium-temperature range), and (3) negative formation energy (a preliminary proxy for stability). We select ternary and quaternary crystals of ≤40 sites having energy above hull <0.08 eV/atom in Materials Project as our dataset (training + test) for ground-state properties, *i.e.*, $E_g$ and $E_f$, and we use the database from reference[35] as our dataset (training + test) for power factor, where the constant relaxation time approximation under the Boltzmann transport equations (BTE) is used to calculate the TE-relevant labels. The final dataset has 34,784 crystal structures with ground-state properties ($E_g$, $E_f$) from Materials Project. Only 4,284 crystals have corresponding calculated power factor labels from reference.[35]

To tackle the sparse-label problem of the excited-state property, we train a semi-supervised VAE.[36] The semi-supervised VAE allows us to jointly train the dataset with full entries of ground-state properties and partial entries of calculated power factor. The semi-supervised VAE developed in this design case consists of a normal VAE, a target-learning branch that maps its entire latent space to ground-state property labels, and a subset of the latent space to the TE property, power factor. The property-mapping loss of the semi-supervised VAE includes one more component compared with the previous VAE in case 2, *i.e.*, the regression for calculated power factor of 4,284 crystals.

Two crystals out of a total 28 (7.1%) designed crystals are shown in Figure 6. After BoltzTraP calculation, the two design crystals are found to have state-of-the-art power factors, comparable with the best TE materials (for example, GeTe[37] has a similar *c*-axis power factor value to the two FTCP-generated candidate materials). In Figure 6, the doping level and temperature are treated as user inputs. In this design case, the following domain-knowledge-based inverse design criteria are selected: (1) a power factor as large as possible, (2) a bandgap between 0.3 and 1.5 eV (desirable for low- and medium-temperature range), and (3) negative formation energy (a preliminary proxy for stability). We sample the latent space to generate 28 unique crystals after filtering out compositions that exist in the database (compositional uniqueness). The median dissimilarity value of designed crystals is 0.67, and five out of 12 valid crystals have a dissimilarity value above 0.75 (structural uniqueness) shown in section S4.3. (The two designed crystals have dissimilarity values of 0.80 and 0.53.) To examine the designed crystals, we conduct structural relaxation using DFT to obtain the final atomic coordinates and perform BoltzTraP calculations to obtain power factor values.[31]

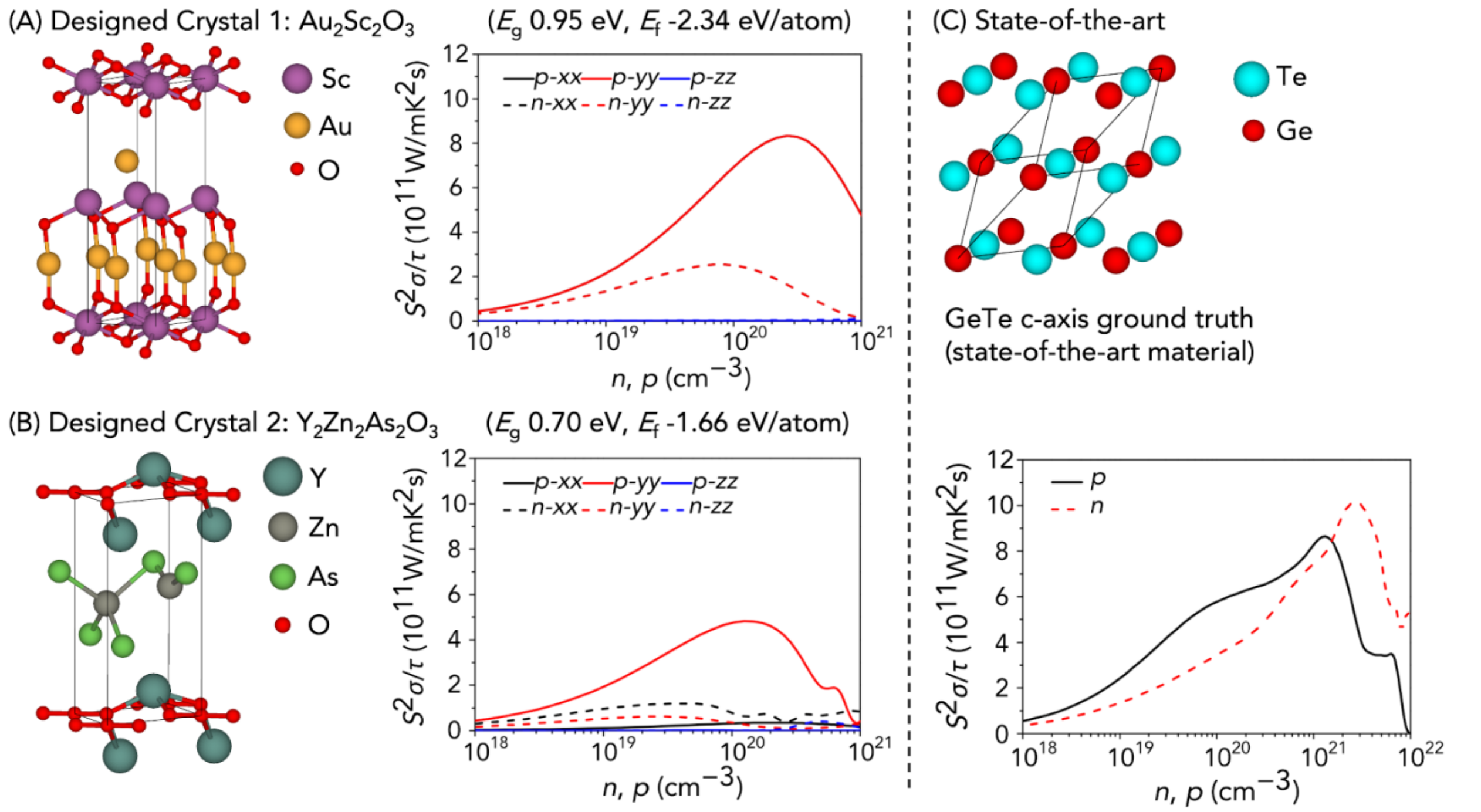

Figure 6. Power factor comparison between two FTCP-designed crystals and the state-of-the-art

(A and B) (A) and (B) show two FTCP-designed crystals with large TE power factors. The two designed crystals are calculated to have comparable power factor (*c*-axis only) with the state-of-the-art high-performance TE material, cubic GeTe, shown in (C). The composition of the two designed crystals do not exist in the Materials Project database (compositionally unique). Their simulated power factor values (divided by the relaxation time) are plotted as a function of doping level at room temperature, both for *n* and *p* doping along the *x*, *y*, and *z* crystal directions in (A) and (B). While $E_f$ and $E_g$ validations were performed using DFT (after structural relaxation), all power factor values were computed under a constant relaxation time approximation using the BoltzTraP package.

## Opportunities for improvement

### Synthesizability challenge

The ultimate goal of inverse design is to pass the designed crystal, supposedly possessing the target property, to an experimental setting, and synthesize the designed crystal. However, the translation from a generative-model-based inverse design algorithm to an actual experimental synthesis of a designed inorganic crystal remains to be demonstrated in the field. More broadly, the topic of synthesis prediction has motivated large research centers (*e.g.*, Center for Next Generation of Materials Design, at https://www.cngmd-efrc.org) indicating both the importance of the topic and the fact that much work remains to be done.

In most prior art to date, instead of experimental validation, proposed inverse design algorithms proffer stability validation using either first-principles calculations,[5,6,13,19-21] or ML surrogate models.[20,22] Among these theoretical stability validations, the majority use negative (or small positive) formation energy as a preliminary proxy for stability (metastability)[6,13,21-23] or small energy above hull as another proxy in the structure-varying cases (limited to specific chemical systems, *e.g.*, $V_xO_y$).[13,19] The authors also adopt this preliminary proxy of negative formation energy in the design cases (as a design criterion) and ensure the dataset used having a small energy above hull, 0.08 eV/atom. (For FTCP, being both composition and structure varying, using energy above hull poses a challenge because hull diagrams of many chemical systems need to be queried, and especially for quaternary chemical systems, hull diagrams are absent for many chemical systems.) However, a negative formation energy design target, and a small energy above hull in the dataset, are not enough to guarantee synthesizability. In fact, the authors attempted to experimentally validate our model by synthesizing the FTCP-designed $Mn_2Co_2Si_5$ in case 1 (DFT-calculated $E_f = -0.326$ eV/atom for $-0.5$ eV/atom target, selected based on synthesis considerations compared with other designed crystals), but upon experimental exploration using an arc melting furnace from pure metal powders, this compound was revealed to decompose into sub-species. A DFT calculation further validates this observation by yielding a decomposition energy of 177 meV/atom.

After this failed experimental validation, the authors realize the importance of optimizing for "precision" (and not "recall" or "F1 score"), framed in language often used in classification tasks. While the FTCP algorithm can design tens of thousands of candidates per hour, experimental throughput is finite. As such, there is a high penalty for false positives. We only need one successful candidate to be experimentally accessible, to succeed at inverse design, but we want to avoid wasting experimental throughput on unsuccessful candidates.

As a potential pathway to future experimental validation, we explore adding a naive synthesizability metric (proxy) named "ICSD score," indicating whether there is an entry for a material in the ICSD. This metric is one method Jang *et al.* have previously explored[25,26] and often used in drug discovery.[38] The metric is obtained by cross-referencing ICSD and Materials Project databases; those compounds in the Materials Project database with an ICSD entry are labeled "synthesizable (1)," and the rest "not synthesizable (0)." To accommodate the synthesizability metric, we add an output to the target-learning branch to map the latent space to ICSD score. By including the ICSD score loss in the overall property-mapping loss, the latent space organizes to reflect the gradient of both the user-specified target properties and the ICSD score illustrated in Figure 7. During inverse design, although the ICSD score intrinsically contains many false negatives (synthesizable crystals not having ICSD entries), the ICSD score still provides us with information on where to sample in addition to the target properties.[39-43]

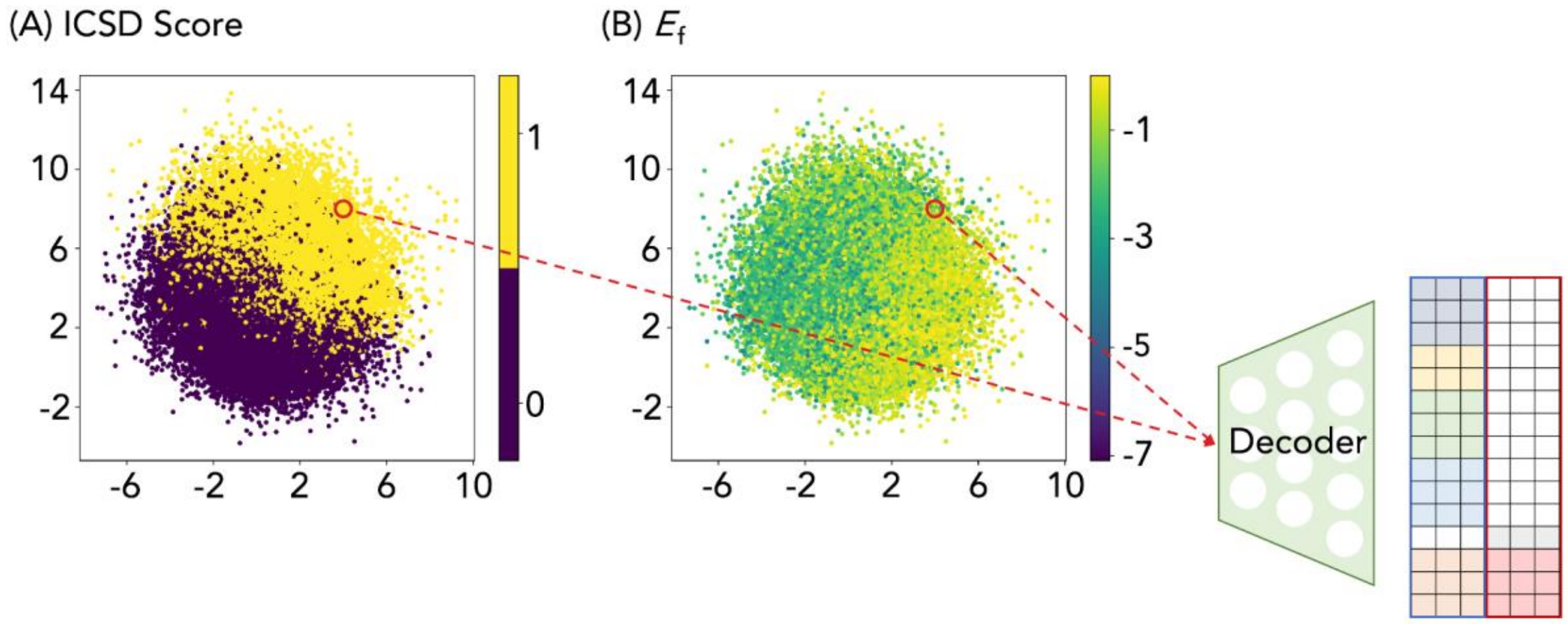

Figure 7. Property-structure latent space with the addition of ICSD score

(A and B) (A) and (B) shows the property-structured latent space colored by the actual ICSD score, and the actual formation energy, respectively. Visualization of the property-structured latent space is with the first and the third dimension (256 dimensions in total) of case 1 as an example, after trained with the target-learning branch having an additional output to predict ICSD score. ICSD score is a naive synthesizability metric predicting whether a point in the latent space would have an ICSD entry. Crystals having ICSD entries are labeled 1, indicating "synthesizable", and the rest 0, indicating "not synthesizable." The latent-space colored points shown are crystals in the training set. From the latent space coloring, a separation between the existence and the absence of ICSD entries in (A) is fulfilled on top of the previous $E_f$ property gradient in (B). This benefits the sampling by letting the user sample from the region where the crystals are of higher likelihood to be synthesized (*i.e.*, ICSD scores of 1).

The addition of the synthesizability metric also demonstrates the flexibility of the framework to add different metrics during inverse design, should consensus on synthesizability or stability metrics be agreed upon by the field. Besides placing the prediction of the additional metric as an additional output of the target-learning branch, placing a classification model on the designed crystals (after decoding), *i.e.*, filtering the designed crystals instead of the latent points, is also a feasible way. As mentioned previously, much work remains to be completed to develop widely accepted synthesizability metrics. Using the examples presented in this study as guides, the FTCP framework should be able to accommodate future metrics.

## Invariance challenge

An aspect of the crystallographic representation that is often considered, arising from the property-prediction applications, is invariance. A crystallographic representation should satisfy full invariances, including translational, rotational, permutational (if there is an order of sites/lattice parameters in the representation), and supercell invariances, because these invariances guarantee different descriptions (such as different CIFs) of the same crystal having the exact same representation, which always results in the same value of the predicted property. A notable class of property-prediction (noninvertible) representation that satisfies full invariances is graph-based, *e.g.*, CGCNN,[27] MEGNet,[1] and SchNet.[44]

However, generative-model-based inverse design (not directed evolution) algorithms need to generate the exact crystal structure (CIF) as the output. It is comparably easy to translate the crystal structure to an invariant representation, but the backward translation from the invariant representation to the exact crystal structure is difficult to implement given this is a one-to-many problem. In prior art, we observe that most of the invertible representations do not satisfy any invariances (Table S1) except for the following two (we exclude the composition-only representation[22] from invariance discussion because there is no geometry involved):

1. Crystal site feature embedding (CSFE)[21] satisfies translation invariance. CSFE assumes a 3D grid representation for perovskite structures and fills in the grid with various site/elemental features, which in essence is a composition representation formatted by structure. Since nothing geometrical except for elemental arrangement is included in CSFE, it satisfies translational invariance.
2. Concatenated spectrum representation of composition and powder XRD pattern[23] satisfies full invariances, but it is not fully invertible, *i.e.*, only outputting a powder XRD pattern, which is hard to convert to a crystal unit cell with algorithmic automation.

Therefore, to our knowledge, there are currently no invertible crystallographic representations that satisfy full invariances. Satisfying full/partial invariances for a generalized representation (both composition and structure varying) is an even harder task. This is an interesting and open field of research.

For the current FTCP representation, it also does not satisfy any invariances, determined by the real-space features that solely guarantee invertibility. Kim *et al.*[19] have shown that performing data augmentation for the real-space representations will lead to a more balanced generation of crystal structures for Mg-Mn-O systems. Given a much larger chemical range ($10^5$–$10^6$) is considered in this study, data augmentation is not feasible. The current formulation of reciprocal-space features only preserves permutational invariance to the order of site inputted (atom indexing), but it has the potential to be formulated to preserve more invariances given the invariances in the powder XRD pattern. To provide some quantification, although the FTCP representation is mainly developed to do inverse design, we present the performance degradation used for property prediction/mapping due to translation, rotation, permutation, and different supercells in section S5.

## Conclusion and future work

We present a framework for general (both composition- and structure-varying) inverse design of inorganic crystals, called FTCP. This method uses an invertible crystallographic representation comprising concatenated real- and inverse-space features of crystals, where the real-space features are CIF-like, guaranteeing invertibility, and the reciprocal-space features embed periodicity and convoluted elemental properties. By jointly training a VAE with a feedforward target-learning branch, we obtain a probabilistic property-structured latent space that allows for inverse design of crystals with user-specified properties through sampling, decoding, and postprocessing. We use FTCP to perform inverse design and design unique solid-state materials with targeted $E_f$, $E_g$, and TE power factor with various chemistries and crystal structures. It is noteworthy that the last property, power factor, is an excited-state property that is challenging to calculate from first principles, yet remains accessible using our inverse design approach through semi-supervised learning. We demonstrate that FTCP can design new crystalline materials that are not in the training set and are dissimilar from known structures. We validate these designed crystals using DFT structural relaxation and confirm their properties by DFT and BoltzTraP calculations. FTCP achieves improvement over random (the probability of finding a material with the user-specified target property by randomly picking from the dataset) ranging from 38.8% to 560%. As a possible pathway to an experimental validation, we explore the addition of a naive synthesizability metric, the existence of an ICSD entry, to further address the synthesizability challenge and demonstrate that FTCP has the flexibility to simultaneously consider this additional metric alongside user-specified target properties. We also comment on the invariance challenge faced by structure-conscious invertible crystallographic representations, including the FTCP representation.

# Experimental procedures

## Resource availability

### Lead contact

Further information and requests for resources should be directed to and will be fulfilled by the lead contact, Tonio Buonassisi (buonassi@mit.edu).

### Materials availability

This study did not generate new unique reagents.

### Data and code availability

The BTE-calculated power factor is from reference,[35] and the rest of the dataset are queried from the Materials Project[9] in November 2019. (Note a query with the same criteria now would yield a different number of crystals from the recorded number in the study due to the updates and the addition of crystals of the Materials Project.) Source codes, and trained parameters, are available at https://github.com/PV-Lab/FTCP.

## Real-Space Features

To effectively represent a crystalline unit cell in the real space, we extract the necessary information in the CIF, and we concatenate the following information matrices to form CIF-like real-space features, as shown in Figure 2A. (Zero padding is used to satisfy the shape specified when necessary.)

- Element matrix of shape ($M$, max($N_{\text{elements}}$, 3)), where $M$ is the length of the one-hot vector representing elements ($M$ = 103 in our case), and $N_{\text{elements}}$ is the largest number of components, *e.g.*, $N_{\text{elements}}$ = 4 when both ternary and quaternary crystals are in the dataset. The number of columns of the element matrix is set to at least three, to conveniently concatenate to the lattice and site coordinate matrices, which have a minimum of three columns.
- Lattice matrix of shape (2, max(3, $N_{\text{elements}}$)). (For the sake of clarity, and without loss of generality, the shape is written as 2 × 3 in section representation.)
- Site coordinate matrix of shape ($N_{\text{sites}}$, max(3, $N_{\text{elements}}$)), where $N_{\text{sites}}$ is the largest number of sites in the unit cell.
- Site occupancy matrix of shape ($N_{\text{sites}}$, max(3, $N_{\text{elements}}$)).

- Elemental property matrix of shape ($K$, max($N_{\text{elements}}$, 3)), where $K$ is the length of the elemental property vector $\boldsymbol{Z}$ ($K$ = 92 from atom_init.json of the CGCNN[27] code).

The concatenated real-space features are of shape ($M + 2 + 2 \cdot N_{\text{sites}} + 1 + K$, max($N_{\text{elements}}$, 3)), where the extra one row of zero padding is added to accommodate the distance of $k$ points in the reciprocal-space features.

## Reciprocal-Space Features

The most common reciprocal-space features of a crystal in materials science are its diffraction pattern. X-ray crystallography is the primary way to study periodic crystals. Modified diffraction images for periodic crystals have been shown to classify their structures accurately.[24] We enrich the information in diffraction images in the reciprocal space by projecting elemental property vector $\boldsymbol{Z}$ to different crystal planes ($hkl$) with an equation modified from discrete Fourier transformation (Equation 1 and S2.2). The authors find mapping $\mathrm{I}_{hkl} = |\mathrm{F}_{hkl}|^2$ to $2\theta$ as in the powder XRD would impose large sparsity in the data, as only a small number of $2\theta$s actually have signals, and the sparsity problem only becomes worse with a finer grid of $2\theta$; thus, authors refrain from this mapping and account for it by prepending the distance of the $k$ point ($hkl$) for each $\mathrm{F}_{hkl}$.

Thus, the reciprocal-space features contain:

- Distance (of $k$ point) matrix of shape (1, $N_{k\text{ points}}$), where $N_{k\text{ points}}$ is the number of nonzero $k$ ($hkl$) points (59 in our case). $N_{k\text{ points}}$ can be treated as a hyperparameter to be tuned. We obtain $N_{k\text{ points}}$ to be 59 by first limiting $|h| + |k| + |l| \leq 3$ ($N_{k\text{ points}}$ = 61) and further tuning ($N_{k\text{ points}}$ reduced to 59).
- FTCP matrix of shape ($K$, $N_{k\text{ points}}$), where column vector $\mathrm{F}_{hkl}$ is arranged according to $hkl$.

After prepending zero padding of shape ($M + 2 + 2 \cdot N_{\text{sites}}$, $N_{k\text{ points}}$), we obtain the reciprocal-space features of shape ($M + 2 + 2 \cdot N_{\text{sites}} + 1 + K$, $N_{k\text{ points}}$).

## Generative model

The generative model (VAE) encodes the FTCP representation into a probabilistic latent space of reduced dimension (256) that can be sampled from. An additional target-learning branch mapping the latent vector to material properties further organizes the latent space to reflect continuous variation of the properties/property gradients (thus the name property-structured latent space). Inverse design of new crystals is achieved by sampling different points other than the existing crystals in the property-structured latent space regions that fulfill the user-defined design targets (enabled by the property gradients). Those sampled points are decoded to FTCP-representation-styled outputs using the decoder. With postprocessing, the CIF is then recovered from the real-space features of the outputs.

We treat the reciprocal-space features as a 1D signal with $N_{k\text{ points}}$ channels. The signal in each channel represents the elemental property projection along a specific ($hkl$) direction. We use a 1D convolutional neural network (CNN) to encode the FTCP representation. The 1D CNN in this work is inspired by PointNet used for 3D point sets classification.[19,45] There is a spatial 1D relationship in the reciprocal-space features. Along this spatial axis, our reciprocal-space representations are arranged according to $k$ (symmetry) points ($hkl$) that are universal in describing the electronic band structure.

The encoder encodes the FTCP representation into a probabilistic normal distribution ($\boldsymbol{z}_{\text{mean}}$ with a diagonal covariance matrix, $\boldsymbol{z}_{\text{variance}}$) using 1D CNN. The decoder, with a symmetrized architecture (using transposed convolutional layers) of the encoder, samples around $\boldsymbol{z}_{\text{mean}}$ with $\boldsymbol{z}_{\text{variance}}$ to reconstruct the FTCP representation, and the reconstruction is regularized by the KL divergence between the latent distribution and the standard Gaussian (zero mean and unit variance) prior. In addition to constraining the latent vector distribution to standard Gaussian prior, we also simultaneously train a feedforward target-learning branch to map $\boldsymbol{z}_{\text{mean}}$ to material properties. The target-learning branch uses Equation 2 where $g$ is fully connected neural networks, $\sigma$ is the sigmoid function (target properties are normalized to the range of 0–1), and $\boldsymbol{z}$ is the latent vector. $\boldsymbol{R}(\boldsymbol{z})$ is the predicted material properties, and in the case of designing for multiple target properties, instead of a scalar, $\boldsymbol{R}(\boldsymbol{z})$ is a vector, of which each entry corresponds to one target material property.

$$\boldsymbol{R}(\boldsymbol{z}) = \sigma\big(g(\boldsymbol{z})\big) \qquad \text{(Equation 2)}$$

In total, we have three losses:

- $L_{\text{reconstruct}}$, the reconstruction loss, using sum of squares between the reconstructed FTCP representation (matrix) and the inputted FTCP representation (matrix). This is summed over a batch.
- $L_{\text{KL}}$, the KL loss, *i.e.*, the KL divergence. This is averaged over a batch.
- $L_{\text{property}}$, the property-mapping loss, using sum of squares between predicted material properties $\boldsymbol{R}(\boldsymbol{z})$ and actual material properties. In case 3, where semi-supervised learning is used, the loss for the incomplete TE label, power factor, is calculated separately also using the sum of squared. (For ICSD score, although a classification task in nature, we still use the sum of squares as loss, taking advantage of the 0–1 range of ICSD score, and the sigmoid activation function.) This is summed over a batch.

The overall loss is

$$L = L_{\text{reconstruct}} + \beta L_{\text{KL}} + \lambda L_{\text{property}} (+\gamma L_{\text{power factor}}) \qquad \text{(Equation 3)}$$

where $\beta$, $\lambda$, and $\gamma$ are user-defined coefficients. (We give the power factor loss a separate coefficient for the semi-supervised learning.) To learn a disentangled representation, we allow heavy penalization of the latent distribution ($\beta > 1$).[46] We minimize the overall loss with a root mean squared propagation (RMSprop) optimizer. We train with a batch size of 256, 200 epochs and a dynamic learning rate. The detailed architecture and hyperparameters of the model used in every design case are tabulated in section S2.4 in supplemental information.

## Acknowledgements

We acknowledge Vladan Stevanovic (Colorado School of Mines) for discussion of applications and validation of FTCP; Shyue Ping Ong (UCSD) for helpful discussions regarding use of ICSD record for stability; Alex Zunger (University of Colorado Boulder) for discussions regarding inverse design; Andy Cooper (Materials Innovation Factory, University of Liverpool) for introducing us to the term "directed evolution" adopted in this manuscript; Aron Walsh (Imperial), Dan Davies (University College London), Keith Butler (Rutherford Appleton Laboratory, UKRI STFC), and Alán Aspuru-Guzik (University of Toronto) for discussions about the upper limits of "stoichiometric inorganic compounds." We appreciate Tian Xie and Xiang Fu (MIT) for the careful review of our manuscript and code. This research is supported by the National Research Foundation, Prime Minister's Office, Singapore under its Campus for Research Excellence and Technological Enterprise (CREATE) program through the Singapore Massachusetts Institute of Technology (MIT) Alliance for Research and Technology's Low Energy Electronic Systems (LEES) research program. F.O., S.S., and Q.Liang acknowledge support from TotalEnergies SE research grant funded through MITei. Y.J. acknowledges the support from the Institute of Information & Communications Technology Planning & Evaluation (IITP) grant funded by the Korea government (No. 2021-0-02068, Artificial Intelligence Innovation Hub). J.L. and X.W. acknowledge support from the Ministry of Education Academic Research Fund R-279-000-532-114,. Y.L. is supported by the National Key Research and Development Program of China (Grant Nos. 2017YFB0702901 and 2017YFB0701502) and the National Natural Science Foundation of China (Grant No. 91641128). S.J. and K.H. acknowledge funding from the Accelerated Materials Development for Manufacturing Program at A∗STAR via the AME Programmatic Fund by the Agency for Science, Technology and Research under Grant No. A1898b0043. G.X. is grateful for the support by the Scientific Computing and Data Analysis section of the Research Support Division at Okinawa Institute of Science and Technology Graduate University (OIST). Q.Li is supported by the National Research Foundation (NRF) fellowship grant NRFF13-2021-0106. A.G.A. acknowledges support from Solar Energy Research Institute of Singapore (SERIS). SERIS is a research institute at the National University of Singapore (NUS). SERIS is supported by the National University of Singapore (NUS), the National Research Foundation Singapore (NRF), the Energy Market Authority of Singapore (EMA), and the Singapore Economic Development Board (EDB).

## Author contributions

Z.R., K.H., and T.B. conceived of this study. Z.R. and S.I.P.T. designed the algorithm and developed and tested the crystallographic representation and the VAE. J.N., G.X., K.H., and Y.J. performed validation tests using DFT and BoltzTraP. J.N. contributed to the analysis of sources of error and role of crystal structure relaxation in correcting for these. F.O. and J.L. contributed to the discussion on the DFT validation and the gap between DFT and experiment. J.L. performed the experiment. S.S., R.Z., and Q.Liang contributed to the development of the synthesizability metric. X.W. and J.L. provided important contributions to improve the neural network (VAE) structure. Q.Li and S.J. provided important intellectual contributions to the section on invariances of crystal representation and shaped the ablation study. Y.L. performed experimental validation and further DFT confirmation on the designed $Mn_2Co_2Si_5$ alloys. A.G.A., Y.L., X.W., Q.Li, S.J., K.H., Y.J., and T.B. supervised different elements of this research. Each co-author contributed portions of the manuscript, the writing of which was coordinated by Z.R., S.I.P.T., S.J., K.H., and T.B. All co-authors approved the final version of the manuscript.

## Declaration of interests

Some of the authors (Z.R., K.H., T.B.) have filed a disclosure on algorithms related to FTCP. Some of the authors (Z.R., K.H., T.B.) are founders and shareholders of a start-up Xinterra, designed to accelerate the development of materials using ML methods.

# Supplemental Experimental Procedures

## S1. Prior Art on Invertible Crystallographic Representations

Table S1. Invertible Crystallographic Representations Used in Generative-Model-Based Inverse Design Algorithms, Related to Figure 1

| Crystallographic Representation | Algorithms using the Representation | Composition-Varying | Structure-Varying | Invertible | Invariances Satisfied* |
|---|---|---|---|---|---|
| In This Study | | | | | |
| Fourier-Transformed Crystal Properties (FTCP) Representation | FTCP‡ | Yes | Yes | Yes | NIL |
| Prior Art: Composition-Only Representations | | | | | |
| Composition Encoding | MatGAN[1†] | Yes | N/A | Yes | N/A |
| Prior Art: Structure-Conscious Representations | | | | | |
| Lattice Matrix + Site Fractional Coordinate Matrix(ces) | CrystalGAN[2†], Composition-Conditional Crystal GAN[3†] | No | Yes | Yes | NIL |
| 3D Voxel Descriptor | ZeoGAN[4†] | No | Yes | Yes | NIL |
| 3D Voxel Descriptor encoded to 2D by an autoencoder | iMatGen[5‡], DCGAN[6†] | No | Yes | Yes | NIL |
| Crystal Site Feature Embedding (CSFE) | VAE+target-learning branch[7‡] | Yes | No | Yes | Translational |
| 2D Tensor comprising nonequivalent site coordinates, lattice length, one-hot-encoded space group, and elemental properties | CubicGAN[8†,§] | Yes | Limited to Cubic Structures | Yes | NIL |
| Concatenated Spectrum Representation of | Double VAE + Bayesian Optimization[9‡,§] | Yes | Yes | Limited/No | Full‖ |
| composition and powder XRD pattern | | | | | |

*These are invariances satisfied inherently by materials representation; the invariances achieved by algorithmic implementations are not included, such as rotational invariances achieved by 3D convolutional neural network with augmentation[10].

†Algorithms using generative adversarial network (GAN)

‡Algorithms using variational autoencoder (VAE)

§Came out later than our arXiv preprint

||Full invariances include translational, rotational, permutational (if there is an order of sites/lattice parameters in the representation), and supercell invariances.

# S2. The FTCP Framework

## S2.1 Property-Structured Latent Space

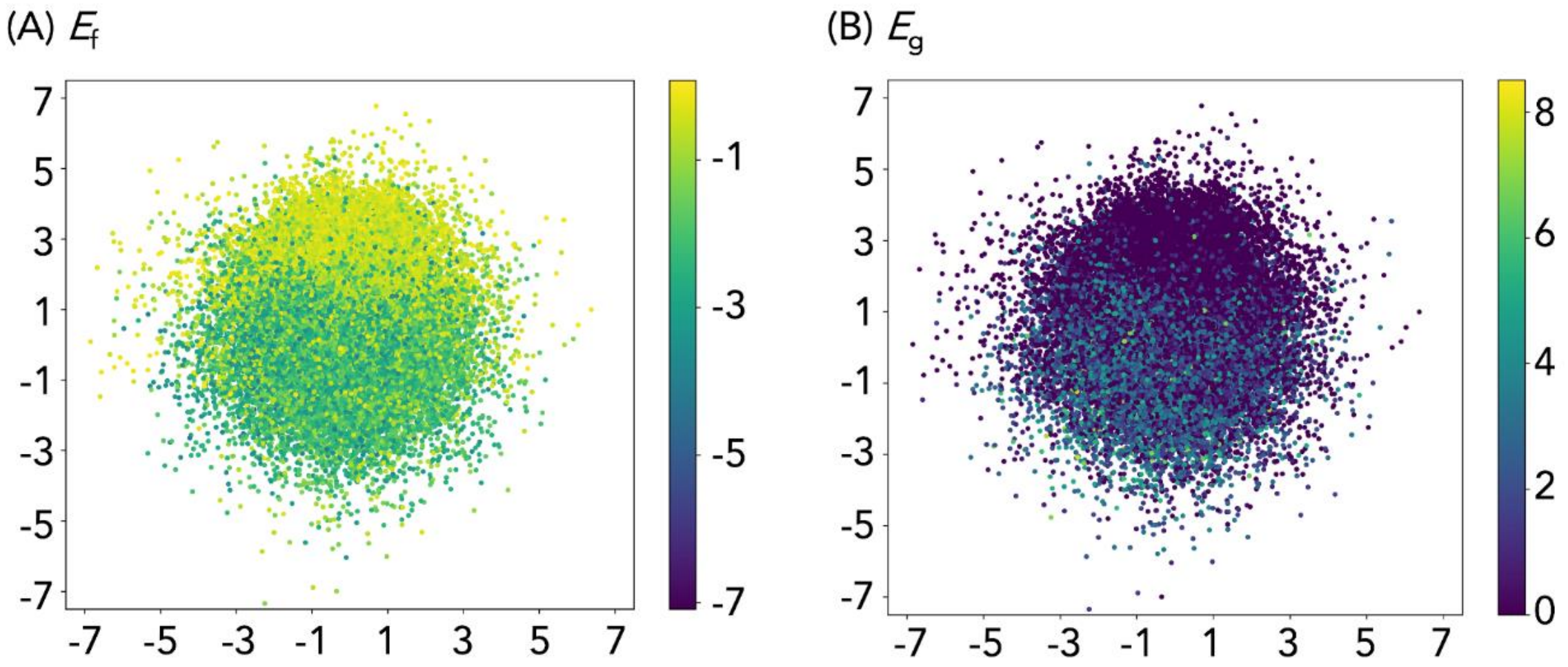

Figure S1. Property-Structured Latent Space, Related to Figure 2 and 7.

Visualization of the property-structured latent space with the third and the 12th dimension (256 dimensions in total) of Case 2: designing for $E_g$ and $E_f$ together in Section *Case 1: Design for Formation Energy and Case 2: Design for Bandgap (with formation-energy constraint): Case studies for photovoltaic & optoelectronic applications*. The dots are encoded training data in the latent space, and they are colored by properties: (A) formation energy per atom, and (B) band gap.

We have two observations of the property-structure latent space: (1) the encoded training data are densely packed (forming a near hypersphere), and (2) there is a continuous change of property (property gradient). The dense packing is achieved by minimizing the VAE Kullback-Leibler (KL) loss, and the property gradients are formed by minimizing the property-mapping loss of the target-learning branch (in Section *Model*). In Case 2, discussed in Section *Case 1: Design for Formation Energy and Case 2: Design for Bandgap (with formation-energy constraint): Case studies for photovoltaic & optoelectronic applications*, the target-learning branch is mapping the latent vector to both $E_f$, and $E_g$, and thus the property-structed latent space shows gradients with respect to both $E_f$, and $E_g$.

Our latent space has 256 total dimensions; the choice of which two to show in our figures was a question of data visualization. We choose the first and the third dimensions (out of 256 of the latent vector) to create the scatter plot in Figure 7, and the third and the $12^{th}$ in Figure 2B, and S1, and the $24^{th}$ and the $203^{rd}$ in Figure S2, as specified in the respective figure captions. They are selected because they present clear intuition of high-dimensional property gradients and the dense packing of the latent space similar to a hypersphere. The corresponding properties are noted either in panel titles, *e.g.*, (A) xxx, or in color bar titles.

We don't choose PCA[11], t-SNE[12], or other dimensionality reduction visualization because we want to preserve the near-hyperspherical shape of the latent space through visualization. Visualizing only two dimensions is showing a cross-sectional view of the near hypersphere (as if cut by a plane). Other dimension reduction visualizations, albeit able to preserve the high-dimensional property gradient, lose the near-hyperspherical arrangement of the latent space.

## S2.2 Formulation of Reciprocal-Space Features (Evaluated with Property Mapping)

Table S2. Comparison of Different Formulations of Reciprocal-Space Features

MAE, mean absolute error. All results are calculated with an updated version of Materials Project accessed on 14 Sep 2021. (Materials Project has updated since our design cases, which accessed Materials Project on 22 Jun 2020.) All results are mean values from five-fold cross validation. Dataset used is the same as the one in Case 1, described in Table S4.

| | FTCP with $\mathbf{F}_{hkl}^{1}$* | FTCP with $\mathbf{F}_{hkl}^{2}$† | FTCP with $\mathbf{F}_{hkl}^{3}$‡ |
|---|---|---|---|
| MAE: $E_f$ (eV/atom) | 0.051 | 0.050 | 0.054 |
| MAE: $E_g$ (eV) | 0.204 | 0.216 | 0.244 |

*$\mathbf{F}_{hkl}^{1} = \sum_{i=1}^{N} \boldsymbol{Z_i} \cdot \frac{j}{2}\ln(e^{-j2\pi(hx_i+ky_i+lz_i)}) = \sum_{i=1}^{N} \boldsymbol{Z_i} \cdot \pi(hx_i + ky_i + lz_i)$

†$\mathbf{F}_{hkl}^{2} = \sum_{i=1}^{N} \boldsymbol{Z_i} \cdot (e^{-j2\pi(hx_i+ky_i+lz_i)})^{\frac{j}{2}} = \sum_{i=1}^{N} \boldsymbol{Z_i} \cdot e^{\pi(hx_i+ky_i+lz_i)}$

‡$\mathbf{F}_{hkl}^{3} = \sum_{i=1}^{N} \boldsymbol{Z_i} \cdot Real(e^{-j2\pi(hx_i+ky_i+lz_i)}) = \sum_{i=1}^{N} \boldsymbol{Z_i} \cdot \cos\big(2\pi(hx_i + ky_i + lz_i)\big)$

Because the discrete Fourier transform is a sum of exponentials of imaginary terms, and we need real numbers to make tractable features, we hereby compare three different formulations (preprocessings) in Table S2. Table S2 shows the property mapping mean absolute error (MAE) of the FTCP representation with three different formulations of the reciprocal-space features (in Section *Representation* and Section *Reciprocal-Space Features*). Based on these results, we choose $\mathbf{F}_{hkl}^{1}$ in our study.

## S2.3 Ablation Study of the FTCP Representation

Table S3. Ablation Study of the FTCP Representation

This table is calculated with an updated version of Materials Project accessed on 14 Sep 2021. (Materials Project has updated since our design cases, which accessed Materials Project on 22 Jun 2020.); MAE, mean absolute error; MAPE, mean absolute percentage error; a.u., arbitrary unit. Dataset used is the same as the one in Case 1, described in Table S4.

| | **Real-Space Features** | **Reciprocal-Space Features** | **Real + Reciprocal-Space Features (FTCP Representation)** | **Benchmark** |
|---|---|---|---|---|
| Property Mapping (Encoder + Target-Learning Branch) | | | | |
| MAE*: $E_f$ (eV/atom) | 0.048 | 0.117 | 0.051 | 0.055 (CGCNN) |
| MAE*: $E_g$ (eV) | 0.202 | 0.354 | 0.204 | 0.250 (CGCNN) |
| Reconstruction (Full VAE + Target-Learning Branch) | | | | |
| Accuracy: Constituent Elements (%) | 98.1 | –† | 99.0 | – |
| MAPE: Lattice Constants (*abc*) (%) | 12.5 | –† | 9.01 | – |
| MAPE: Lattice Angles (*αβγ*) (%) | 8.12 | –† | 5.07 | – |
| MAE: Site Fractional Coordinates (a.u.) | 0.047 | –† | 0.045 | – |

*Results are mean values from five-fold cross validation.

†Reciprocal-space features alone cannot reconstruct the 3D crystal, and are excluded in the reconstruction error comparison

We perform the ablation study on the FTCP representation in Table S3 under two scenarios, (1) in property mapping, and (2) in reconstruction. In property mapping, we only use the encoder + target-learning branch (a feed-forward model). We observe that real-space features alone perform the best, while the FTCP representation (real + reciprocal-space features) performs on par with real-space features alone. We also benchmark this property mapping against the state-of-the-art graph representation, crystal graph convolutional neural networks[13] (CGCNN). In reconstruction, we use the full VAE (*i.e.*, encoder, decoder) + target-learning branch. We observe that the FTCP representation achieves higher accuracy in reconstructing constituent elements, and lowers errors in reconstructing lattice parameters, and site fractional coordinates than real-space features alone. (Only using reciprocal-space features cannot reconstruct the 3D crystal, and are thus excluded in this reconstruction comparison.) The overall results of the two scenarios justifies the incorporation of the additional featurizer, reciprocal-space features, along with the real-space features which solely guarantees invertibility.

The incorporation of the reciprocal-space features in the reconstruction has two competing effects, positive by creating more correlation between the input for the reconstruction to latch on (analogous to the fact that predicting for two related tasks help the model learn in multitask learning), and negative by diverting the reconstruction capacity from the real-space features reconstruction. Based on the results, the incorporation has more positive effects in the reconstruction. The validating effect (helping with the reconstruction) of the reciprocal-space features is learnt by model during training, as in multitask learning, instead of explicitly enforced.

Note that performance of CGCNN is slightly worse than reported, such as in the Matbench study[14]. In this benchmark, we use the same hyperparameters as in the Matbench study, and the difference in performance can be attributed to the difference in datasets used. The Matbench study uses the whole Materials Project, while this benchmark only uses ternary compounds with $\leq 20$ sites, and $< 0.08$ eV/atom energy above hull as used in Case 1. We also note, just slight changing our dataset to $\leq 0.1$ eV/atom energy above hull, CGCNN performs better, achieving 0.028 eV/atom (MAE for $E_f$), and 0.190 eV (MAE for $E_g$).

## S2.4 Datasets, Architecture and Hyperparameters

Table S4. Datasets, Architecture, and Hyperparameters Used in the Design Cases

| | Case 1 | Case 2 | Case 3 |
|---|---|---|---|
| Datasets | | | |
| Number of Components | Ternary | Ternary, and quaternary | Ternary, and quaternary |
| Number of sites | $\leq 20$ | $\leq 40$ | $\leq 40$ |
| Energy above hull (eV/atom) | $< 0.08$ | $< 0.08$ | $< 0.08$ |
| Train Test Split | 80% (training set), and 20% (test set) | | |
| Architecture | | | |
| Encoder | The encoder uses three 1D convolutional layers with filter sizes {5,3,3}, number of filters {32,64,128}, and strides {2,2,1}. After every convolutional layer, batch normalization is immediately applied. Leaky-ReLU activation function with parameter 0.2 is used after every batch normalization. The output of the convolutional layers is transformed to a 256-dimensional latent vector $z$ via two fully connected layers of 1024 nodes with sigmoid activation function, and of 256 nodes with linear activation function. | | |
| Decoder | The 256-dimensional latent vector is first transformed to the output shape of the last convolutional layer in the encoder via a fully connected layer with ReLU activation, and a reshape layer. After applying batch normalization to the transformed output, three 1D transposed convolutional layers are used with filter sizes {3,3,5}, number of filters {128,64,32}, and strides {1,2,2}. Batch normalization and ReLU activation function are used between the transposed convolutional layers. The last layer of transposed convolutional layer is followed by sigmoid activation function (corresponding to reconstructed input of the VAE). | | |
| Target-Learning Branch | The target-learning branch connects to the latent space, and takes the mean of the 256-dimensional latent vector $\boldsymbol{z}_{\mathbf{mean}}$ as input. $\boldsymbol{z}_{\mathbf{mean}}$ is first activated by ReLU function, and then passed through two fully connected hidden layers of 128 nodes, and 32 nodes with ReLU activation functions. Then a fully connected output layer with sigmoid activation function produces a vector of predicted property(ies). The number of nodes of the output layer depends on | | |
| | the number of properties being mapped, *i.e.*, one, two and three for Case 1, 2, and 3. | | |
| Hyperparameters | | | |
| Batch Size | 256 | | |
| Epochs | 200 | | |
| Optimizer | Root Mean Squared propagation (RMSprop) | | |
| Learning Rate | The learning rate is constantly reduced by 0.3 on plateau (with a minimum value of 1e-6), while the training loss has not improved over four epochs. | | |
| | Epoch 0: 5e-4<br>Epoch 50: 1e-4<br>Epoch 100: 5e-5 | Epoch 0: 5e-4<br>Epoch 50: 1e-4<br>Epoch 100: 5e-5 | Epoch 0: 8e-4<br>Epoch 50: 2e-4<br>Epoch 100: 5e-5 |
| β | 2 | 2 | 3 |
| λ | 10 | 10 | 20 |
| γ | – | – | 5 |
| Scale (for the scaled unit Gaussian noise used in the local perturbation sampling) | 0.4 | 0.6 | 0.9 |

For hyperparameters, batch size is tuned between 32 and 1024 by randomly sampling various powers of 2; learning rate is tuned with a resolution of a logarithmic scale; β, λ, and γ, are randomly sampled from grids, {1, 2, 3, 5, 10, 100}, {1, 10, 50, 100}, and {5, 50, 200, 500}. The scale for the scaled unit Gaussian noise used in the local perturbation sampling ranges from 0.3 to 3, and "is determined through trial-and-error balancing of exploration and exploitation, bounded by structure design errors (for lattice) of <20% while still ensuring maximum possible exploration", discussed in greater detail in Section *S2.5 Comparison of Sampling Strategies*. Users of FTCP are encouraged to do their own hyperparameter tuning as the dataset is bound to be different. (These hyperparameters are from our design cases that accessed the Materials Project on 14 Sep 2021, and since then, the Materials Project has been updated.)

## S2.5 Comparison of Sampling Strategies

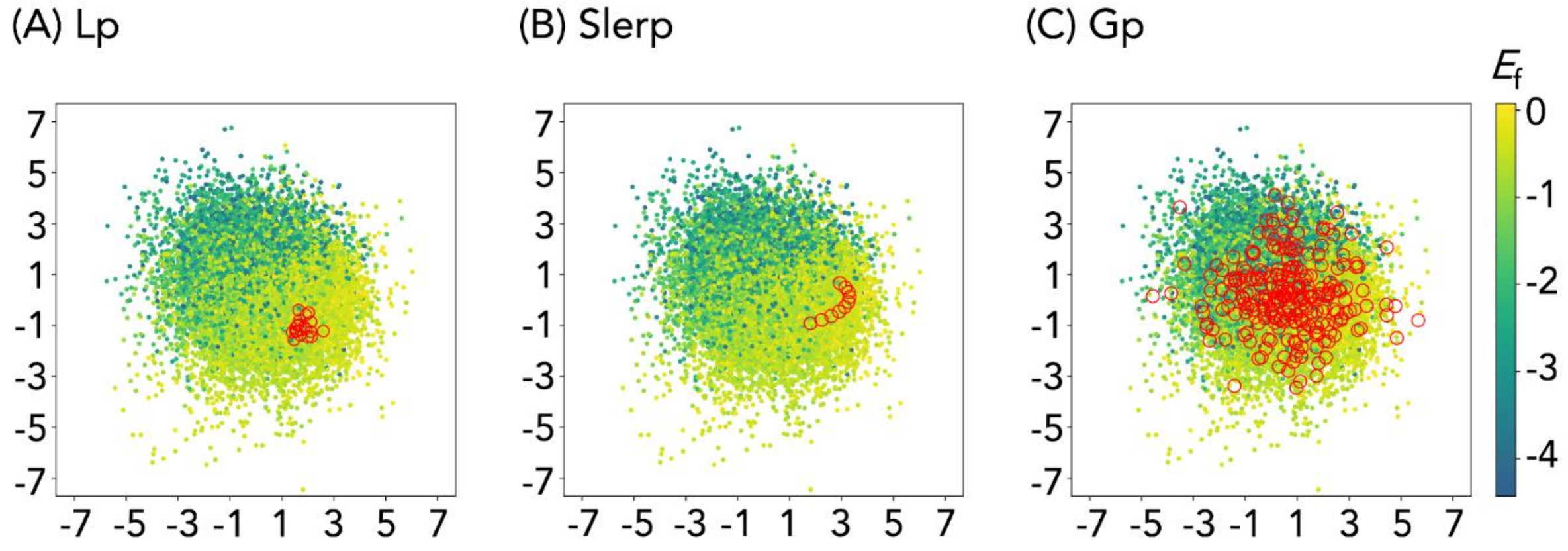

Figure S2. Three Sampling Strategies

(A) local perturbation (Lp), (B) spherical linear interpolation (Slerp), and (C) global perturbation (Gp) in the latent space. The latent space is visualized with the $24^{th}$ and the $203^{rd}$ dimension (256 dimensions in total) of Case 1. Colored latent points are encoded training data. Red circles indicate sampled latent points.

In Figure S2, we illustrate three sampling strategies we explore, namely local perturbation (Lp), spherical linear interpolation[15] (Slerp), and global perturbation (Gp). Lp behaves similar to the decoder, sampling (perturbing) around a latent point with a scaled unit Gaussian noise. Figure S2A shows the sampling around one latent point. Slerp, as the name suggests, sampling with spherical linear interpolation between a pair of latent points. Figure S2B shows the sampling between one pair of latent points. Gp samples using a Gaussian with the mean, and the variance of the latent space, shown in Figure S2C. In addition to the decoder-like Lp, Slerp is good for sampling in between (interpolation) points in the training set, and Gp samples with high overlapping with the latent space.

As a demonstration, we evaluate the performance of three sampling strategies in Case 1, where the design target is $E_f$ = -0.5 eV/atom. For Lp, we prioritize sampling around training data that have the closest $E_f$ values (with scale = 0.4 for the scaled unit Gaussian noise); for Slerp, we prioritize sampling in between training data that have the closest $E_f$ values (with 10 interpolations between each pair); for Gp, as it cannot do property-driven sampling, we simply sample with the mean and the variance of the latent space. We use a metric gauging the structure design errors of rediscovered crystals. The rediscovery refers to rediscovering crystals in the test set (20% of the dataset) via sampling. For a fair comparison, we use the same number (500) of rediscovered crystals in the test set for all strategies in calculating the structure design errors. The structure design errors are between the decoded FTCP representation of rediscovered crystals via sampling (without DFT structural relaxation), and the FTCP representation of the actual corresponding crystals in the test set. For structure design errors, we look at the mean absolute percentage error (MAPE) of designed lattice constants, and designed lattice angles, and the mean absolute error (MAE) of designed site fractional coordinates (the change from MAPE to MAE is due to zero values of site fractional coordinates leading to nonsensical percentage errors). The results are tabulated in Table S5.

Table S5. Evaluation of Three Sampling Strategies in terms of the Structure Design Errors of Rediscovered Crystals

The structure design errors are calculated for 500 rediscovered crystals in the test set from every sampling strategy. The designed crystals here do not go through DFT structural relaxation, but decoding from sampled latent points, and postprocessing. Lp: local perturbation, Slerp: spherical linear interpolation, Gp: global perturbation. All results are calculated with an updated version of Materials Project accessed on 14 Sep 2021. (Materials Project has updated since our design cases, which accessed Materials Project on 22 Jun 2020.) MAPE: mean absolute percentage error, MAE: mean absolute error, a.u.: arbitrary unit. Dataset used is the same as the one in Case 1, described in Table S4.

| Structure Design Errors of Rediscovered Crystals | Lp | Slerp | Gp |
|---|---|---|---|
| MAPE:<br>Designed Lattice Constants (*abc*)<br>(%) | 16.1 | 21.9 | 28.5 |
| MAPE:<br>Designed Lattice Angles ($\alpha\beta\gamma$)<br>(%) | 11.5 | 16.7 | 30.6 |
| MAE:<br>Designed Site Fractional Coordinates<br>(a.u.) | 0.28 | 0.29 | 0.30 |

We find that new crystals obtained by Lp mainly experience elemental substitution compared to its root crystal. This agrees well with the prevailing design principles for new crystalline materials, which makes use of manual substitution of certain elements in the unit cell[16]. Slerp and Gp generate samples that experience more structural change; however, the structure design errors in the rediscovered crystals are much higher. Overall, we observe this tradeoff between exploitation, and exploration when sampling close to, or far from the latent point of known crystals (*i.e.*, in the training set).

To understand more about the results, we review the training process of VAE: the encoder maps the input to $\boldsymbol{z}_{\mathbf{mean}}$ and $\boldsymbol{z}_{\mathbf{variance}}$, and the decoder maps the vicinity of $\boldsymbol{z}_{\mathbf{mean}}$ according to $\boldsymbol{z}_{\mathbf{variance}}$ back to the input itself; the Kullback-Leibler (KL) loss encourages $\boldsymbol{z}_{\mathbf{mean}}$ and $\boldsymbol{z}_{\mathbf{variance}}$ to follow those of a unit Gaussian (prior distribution of the latent variable $\boldsymbol{z}$), while the reconstruction loss encourages clustering of latent points of similar inputs, and the property-mapping loss promotes clustering of latent points of similar properties; in the joint effect of various losses, $z_{mean}$ is densely packed around the center of latent space, while latent points of similar input and similar properties are near one another. This dense-packing and similarity-clustering enables “interpolation” from one known latent point to another yielding new crystals, thus achieving embedding the input to a continuous latent space. Due to the interpolation nature of the formed latent space, the validity distribution is not uniform across the latent space. Instead, high validity clusters around latent points of known crystals (because of how decoder works), and lower in between.

When a sampling strategy does a higher degree of exploitation (sampling close to latent points of known crystals), such as Lp, the new sampled crystals show greater resemblance to known crystals, and the designed structures show higher validity, *i.e.*, lower structure design errors. On the other hand, when a sampling strategy performs a higher degree of exploration (sampling far away from latent points of known crystals), such as Slerp, and Gp, the new sampled crystals show higher variety, and the designed structures show lower validity, *i.e.*, higher structure design errors. (Although for Lp, we can control the exploitation-exploration tradeoff by adjusting the scale for the scaled unit Gaussian noise. In fact, in the design cases, the scale is determined through trial-and-error balancing of exploration and exploitation, bounded by structure design errors (for lattice) of <20% while still ensuring maximum possible exploration.

Given the long development cycle for solid-state materials[17], it is of critical importance to have an accurate initial guess. The density functional theory (DFT) calculations to validate FTCP-designed crystals are computationally expensive. DFT calculations conducted in this study takes around 1-2 hours per crystal with 16 CPU cores. More complex DFT calculations, such as defect calculations, can take days to complete[18]. Given all the above considerations, we choose Lp in our three design cases, which gives us the lowest structure design, as our property-driven sampling strategy.

### S2.6 Sources of Error for Validity Rate, and the Need of DFT Structural Relaxation

When designing a new crystal using FTCP, there are two notable sources of error resulting from imperfect elimination of reconstruction loss (*i.e.*, misplacements of atoms around their lattice sites, and inaccuracy of reconstructed lattice and elements), which contribute to the error of validity rate—(1) decompression error: note that latent space is a reduced-dimension representation of the CIF. As such, there is a decompression error (similar to going from a TIF image to a JPG one), and (2) interpolation error: As mentioned in Section *S2.5 Comparison of Sampling Strategies*, we note that sampling within latent space can generate error. Moving from one point to another in latent space causes the atoms and their positions to shift. Slight changes to a latent-space point, caused by variations in sampling, can also cause the designed crystal to change slightly. From the structure design errors of rediscovered crystals (without relaxation) using Lp in Table S5, we see that FTCP-designed crystals yield crystal parameters very close ($\leq$ 20%) to relaxed ones (in test set calculated by Materials Project). The DFT structural relaxation is like a snap-to-grid. Of course, there are other cases where the decoded crystals are so off (because of interpolation, and the universal acceptability of the postprocessing) that the structural relaxation (with a specific tolerance) would result in unintended crystals or fail altogether.

## S3. FTCP-Designed Crystals, and Their First-Principles Calculations

### S3.1 FTCP-Designed Crystals

Table S6. FTCP-Designed Crystals, Related to Figure 4, 5, and 6.

Only valid structures are listed, namely structurally relaxed crystals. Only composition is listed, and the crystallographic information files (CIFs) can be found at https://github.com/PV-Lab/FTCP. Bolded are target-satisfying crystals.

| Design Cases | Design Targets | FTCP-Designed Crystals |
|---|---|---|
| Case 1 | $E_f$ = -0.5 eV/atom | $Ag_2MgNb$, $LiRu_2Zr$, **$Al_4DyLu$**, **$As_2Li_3Mn$**, $AcAl_2Os$, **$Gd_2In_3Sn_3$**, **$Al_4DyLa$**, $CRbRh_3$, $CuPbSr_2$, $Mn_2Co_2Si_5$, **$Al_4HoPm$**, **$CeFeGe_3$**, $Fe_2LiSe_2$, **$Bi_4Ce_2Cu$** |
| | $E_f$ = -0.3 eV/atom | $CuGaLi_2$, $LiOs_2Si$, $Ca_2LiS$, $BeGe_2Si$, $Ag_2CdN$, $Ag_2AlCa$, **$GaNb_6Pt$**, $GeNb_6$, **$AsNb_6Pt$**, **$Rh_2SbSn$**, CCoLa, $CePb_2Y$, $Rh_2SbSm$, $FeGd_2Ni_3$, $Co_2ErSn$, $Al_2Ge_2U$, $Ge_2U_3$ |
| | $E_f$ = -0.6 eV/atom | NaPdSi, $Ir_5Mg$, $AuCK_3$, $AuK_3N$, **$K_3PbO$**, $K_2RuBr_6$, $KRhO_2$, $Ge_2Rh_2Sc$, $RhScO_2$, $PaSe_4Ti_2$, $CrS_3W_2$, $CuPdF_2$, **AlSrP**, VSrP, MgSbY, MnSbY, $In_4S_8O$, **BaPdSn**, **$CeGeRh_3$**, **$GePrRh_3$**, $RhSm_2Sn_2$, AgAlGd, $AuHo_2Pd$, PdTlTm, SbYbS, $HfPd_2Yb$, $BTh_4$ |
| | $E_f$ = -0.7 eV/atom | $BLi_2Pt$, $CoLi_2Pt$, $As_4BTi_3$, **$Pt_2SnTi$**, $FeRu_2As_2$, $Sr_4SnN_5$, $PtYO_2$, $Fe_4Ge_4Zr_4H$, $Al_6PdZr_{10}$, OsSbS, **$LaPt_5Sb$**, **$As_2PrRu_2$**, PdPm, **$Ge_2Sm_2Zn$**, $CdEuPd_2$, $Ho_2OsSi_2$, **$Ho_2CdGe_2$**, AsErTi, TiTmO, GeYb, $HgYb_2S$, GeLuNa, GeLuMn, **GeLuZr**, $HgHO_2$, $NbRh_2Th$ |
| Case 2 | $E_g$ = 1.5 eV<br>$E_f$ < -1.5 eV/atom | $Be_3KLiO_4$, BaNaSeF, **$RbScS_2$**, $Er_2Zn_2As_2O_3$, $Au_2Sr_2O_5$, $KLiZn_3O_5$, $Bi_3Fe_4LaO_{10}$, CrCuO, $NaSb_2O_5$, CrCuO (different structure), **$Ba_2PdSe_4O_{11}$**, **$LiSmSe_2$**, **$CaFeGaO_4$**, **$Mg_2Tb_4HS_9$**, **$Li_6NaBiO_6$**, **$LiSmSe_2$** (different structure) |
| Case 3 | Power factor as large as possible<br>0.3 eV ≤ $E_g$ ≤ 1.5 eV<br>$E_f$ < 0 eV/atom | **$Au_2Sc_2O_3$**, $Sr_4In_4O_{11}$, $Ba_2Cu_2Te_2F_3$, $Ba_2Ag_2Te_2F_3$, $Ba_6B_2P_2O_7$, $La_2Zn_2As_2O_3$, $YSF_2$, **$Y_2Zn_2As_2O_3$**, $AgBi_2Se_3Cl_2$, $Cs_6Ge_8Au_3$, $Ba_3Sb_2O_2$, $Bi_4S_2O_5$ |

## S3.2 First-Principles Calculations in Case 1, and 2

For all DFT calculations, we performed spin-polarized Perdew-Burke-Ernzerhof (PBE)[19] calculations with projector-augmented wave (PAW)-PBE pseudopotentials[20] as implemented in the plane-wave based *ab initio* package, VASP[21]. We selected pseudopotentials as recommended in Materials Project database (accessed in Nov. 2019).[22] In addition, we performed spin-polarized PBE+$U$ calculations for transition metal (TM) oxides with the $U$-value of TM taken from Materials Project database (i.e. $U$ = 3.7 for Cr).[22] Atomic positions and unit cell parameters are fully relaxed using conjugate gradient descent method with the convergence criteria of 1.0e-5 for energy and 0.05 eV/Å for force with 500 eV cut-off energy. Brillouin zone is used with the k-point densities of 1000 k-points per atoms using the *Pymatgen* package.[23] To compute the formation energy, we adapted GGA/GGA+$U$ mixed approach (see below) with anion correction term adapted in Materials Project (*i.e.*, -1.4046 eV/atom for $O_2$, and -0.6635 eV/atom for S).[24] All band gap values are computed using PBE (+$U$) functional with Blöchl correction-included tetrahedron method for Brillouin zone integration[25] with increased k-point densities (i.e. 1500 k-points per atoms). All GGA+$U$ energies ($E^{\mathrm{GGA+}U}_{\mathrm{Cr,O\ compound}}$) are corrected by Equation S1.

$$E^{\mathrm{GGA+}U\ \mathrm{corr.}}_{\mathrm{Cr,O\ compound}} = E^{\mathrm{GGA+}U}_{\mathrm{Cr,O\ compound}} - n_{\mathrm{Cr}}\Delta E_{\mathrm{Cr}} \quad \text{(Equation S1)}$$

where, $E^{\mathrm{GGA+}U\ \mathrm{corr.}}_{\mathrm{Cr,O\ compound}}$ is the corrected GGA+$U$ energy for Cr, O compound, $n_{\mathrm{Cr}}$ is the number of Cr atoms in the compound, and $\Delta E_{\mathrm{Cr}}$ is the correction energy for Cr atom. In addition, we included an $O_2$ energy correction term (-1.4046 eV/atom) taken from the Materials Project database for the $E_{\mathrm{Cr,O\ compound}}^{\mathrm{GGA}+U}$. After rearranging Equation S1, the correction energy for Cr atom ($\Delta E_{\mathrm{Cr}}$) is then obtained using Equation S2.

$$\Delta E_{\mathrm{Cr}} = (E_{\mathrm{f}}^{\mathrm{calc.}} - E_{\mathrm{f}}^{\mathrm{ref.}})/r_{\mathrm{Cr}} \qquad \text{(Equation S2)}$$

In Equation S2, $E_{\mathrm{f}}^{\mathrm{calc.}}$ is the formation energy calculated in this work, $E_{\mathrm{f}}^{\mathrm{ref.}}$ is the formation energy for the 3-reference structures ($Cr_2O_3$, $Cr_5O_{12}$, and $Cr_6O_{11}$) taken from the Materials Project database[22] all of which are at the convex hull of the Cr-O binary phase diagram, and $r_{\mathrm{Cr}}$ is the fraction of Cr atom. Therefore, $\Delta E_{\mathrm{Cr}}$ can be obtained by calculating the slope of the ($E_{\mathrm{f}}^{\mathrm{calc.}} - E_{\mathrm{f}}^{\mathrm{ref.}}$) vs. $r_{\mathrm{Cr}}$ plot as shown in Figure S3. The adjusted formation energy values are listed in Table S7.

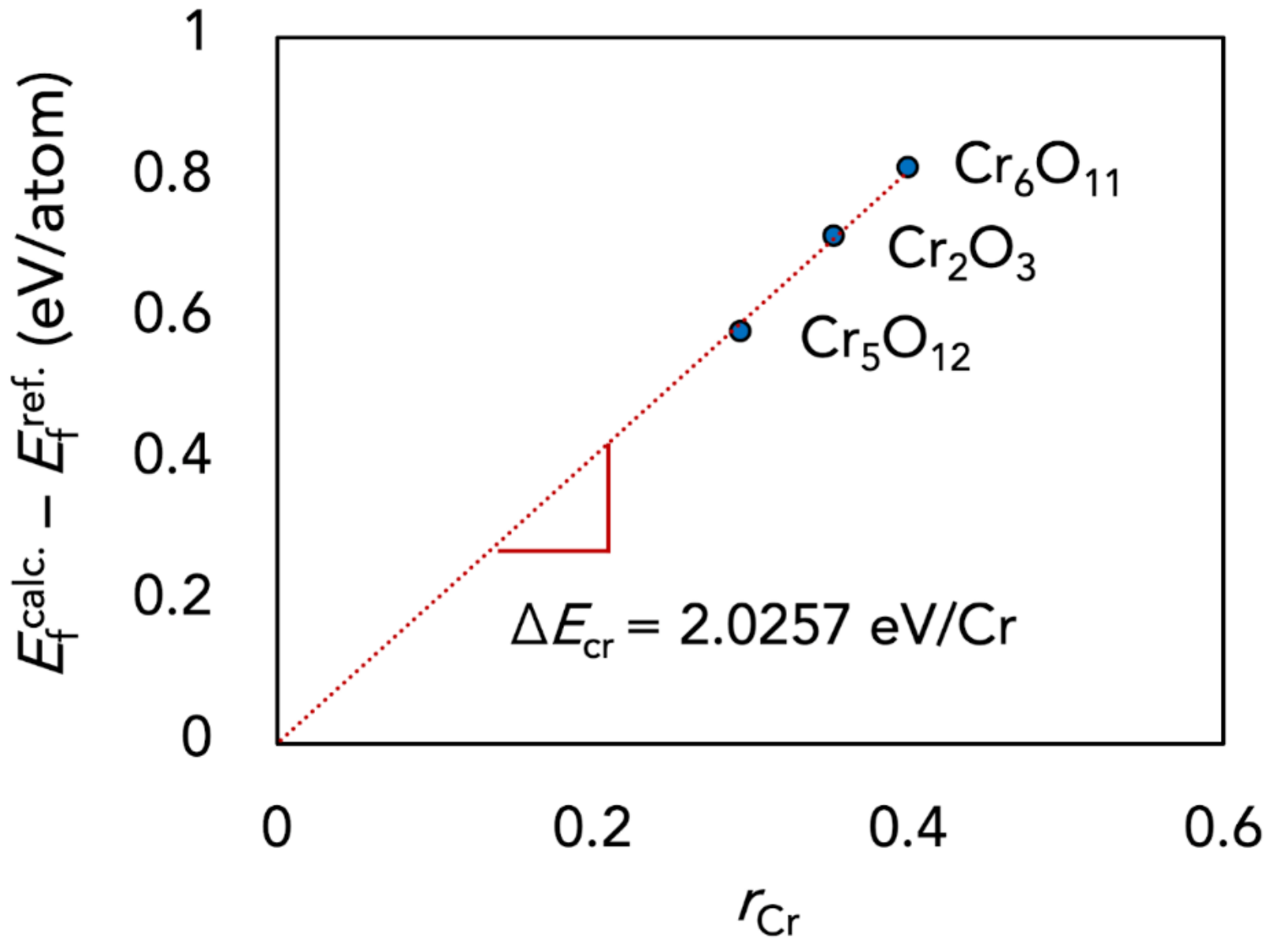

Figure S3. Correction Energy for Cr

Deriving correction energy for Cr ($\Delta E_{\mathrm{Cr}}$) using the GGA/GGA+U-mixed scheme for the 3-$Cr_xO_y$ materials at the convex hull.

Table S7. Results of the Three $Cr_xO_y$ Materials at the Convex Hull After Applying the Correction Energy Term

| Materials | | $E_f$ from MP (eV/atom) | After Correction (eV/atom) |
|---|---|---|---|
| mp-773920 | $Cr_5O_{12}$ | -1.852 | -1.864 |
| mp-19399 | $Cr_2O_3$ | -2.349 | -2.344 |
| mp-1213798 | $Cr_6O_{11}$ | -2.131 | -2.127 |

## S3.3 First-Principles Calculations in Case 3

First-principles calculations of structure relaxations were performed with DFT using the PBE generalized gradient approximation (GGA)[19]. The PAW method[26] as implemented in the VASP code[21] was applied for all the calculations. Energy cutoff of 520 eV for the plan-wave expansion and a Brillouin zone integration spacing of $2\pi \times\ 0.028\ \text{Å}^{-1}$, and $2\pi \times 0.013\ \text{Å}^{-1}$ was used for structure relaxations and total energy calculations, respectively.  We employed the Gaussian method with the width of 0.02 eV to calculate the total energies and band gaps with the convergence criteria of 1.0e-8 eV. The transport coefficients were calculated using the BoltzTraP[27] code with the constant scattering time approximation (CSTA).

# S4. Design Cases

## S4.1 Eighteen FTCP-Designed Crystals for Design Target—$E_f$ = -0.5 eV/atom (Case 1)

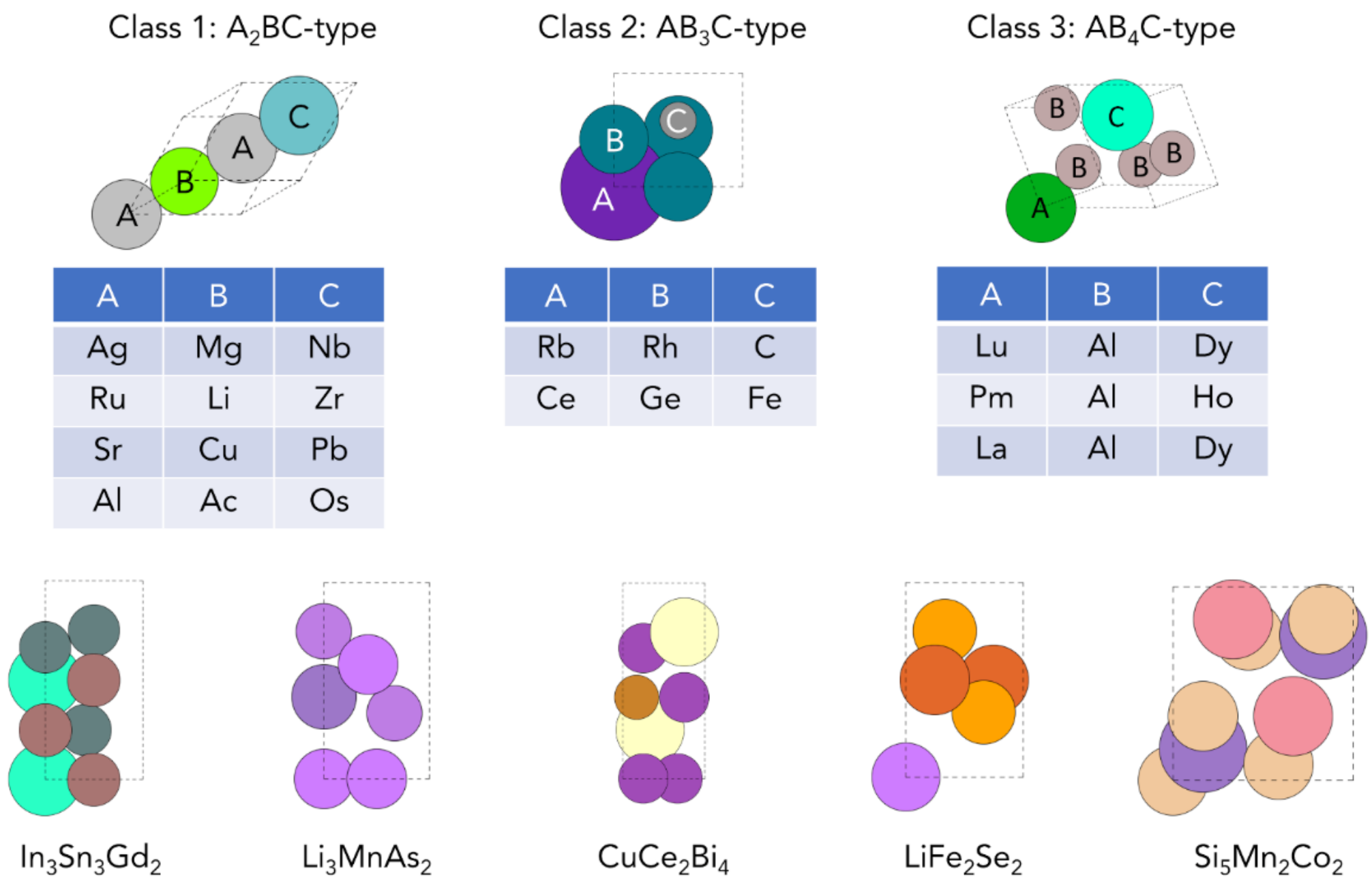

Class 1: $A_2BC$-type

| A | B | C |
|---|---|---|
| Ag | Mg | Nb |
| Ru | Li | Zr |
| Sr | Cu | Pb |
| Al | Ac | Os |

Class 2: $AB_3C$-type

| A | B | C |
|---|---|---|
| Rb | Rh | C |
| Ce | Ge | Fe |

Class 3: $AB_4C$-type

| A | B | C |
|---|---|---|
| Lu | Al | Dy |
| Pm | Al | Ho |
| La | Al | Dy |

Figure S4. Fourteen valid FTCP-Designed Crystals out of a Total of 18, Related to Figure 4.

The designed crystals are unique, *i.e.*, not existent in the Materials Project database. Four invalid (atom-overlapping) variants of $CeFeGe_4$ are not shown here.

Figure S4 shows the FTCP-designed crystal structure of 14 valid crystals (with four invalid ones not shown). There are eight different crystal structures with > 30 different elements in the 14 crystals, showing the FTCP can design crystals accessing a wide range of structures and chemistries.

## S4.2 DFT-Calculated $E_f$ Values of FTCP-Designed Crystals in Case 1

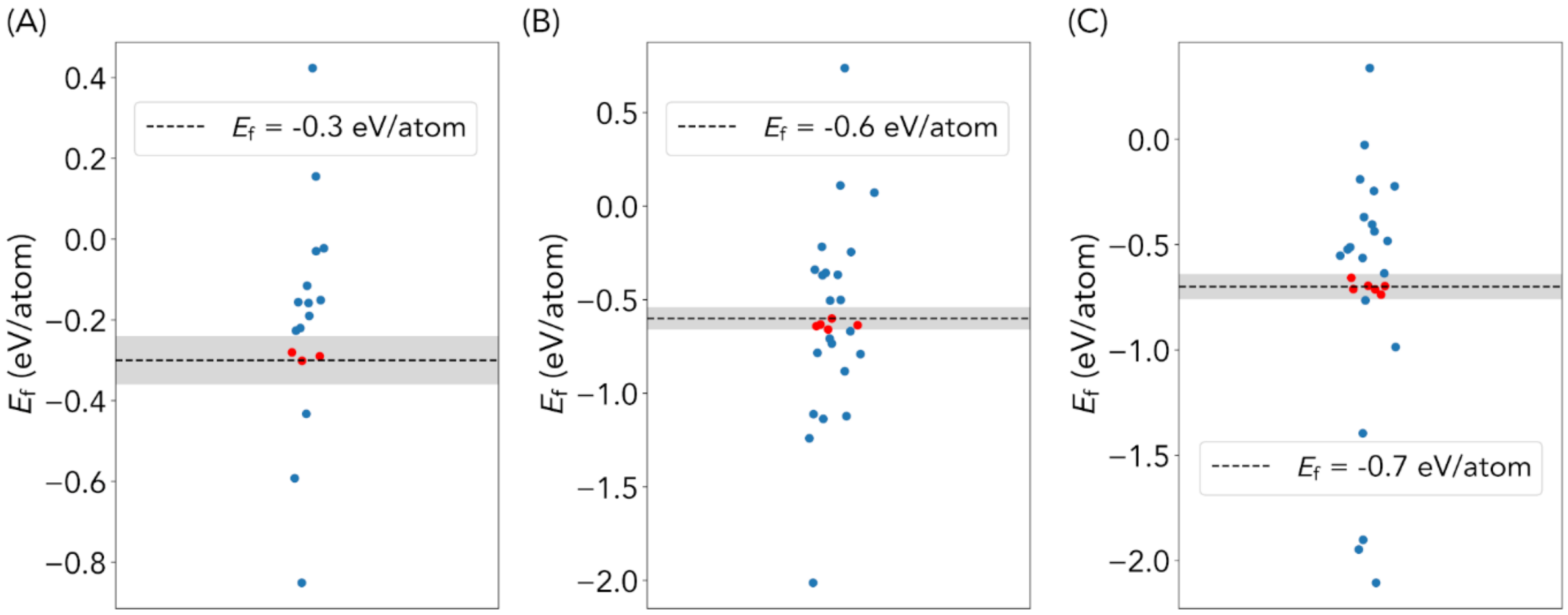

Figure S5. DFT-Calculated Formation Energies of FTCP-Designed Crystals (70 Valid out of a Total of 77), where gray bands indicate target-satisfying regions within tolerance. Related to Figure 5.

With target $E_f$ equal to (A) -0.3, (B) -0.6, and (C) -0.7 eV/atom. Red dots represent target-satisfying designed crystals.

Figure S5 shows three boxplots of DFT-calculated $E_f$ values of FTCP-designed crystals with design target $E_f$ = -0.3, -0.6, and -0.7 eV/atom. The success rates can be found in Table 1.

## S4.3 Dissimilarity Values of FTCP-Designed Crystals

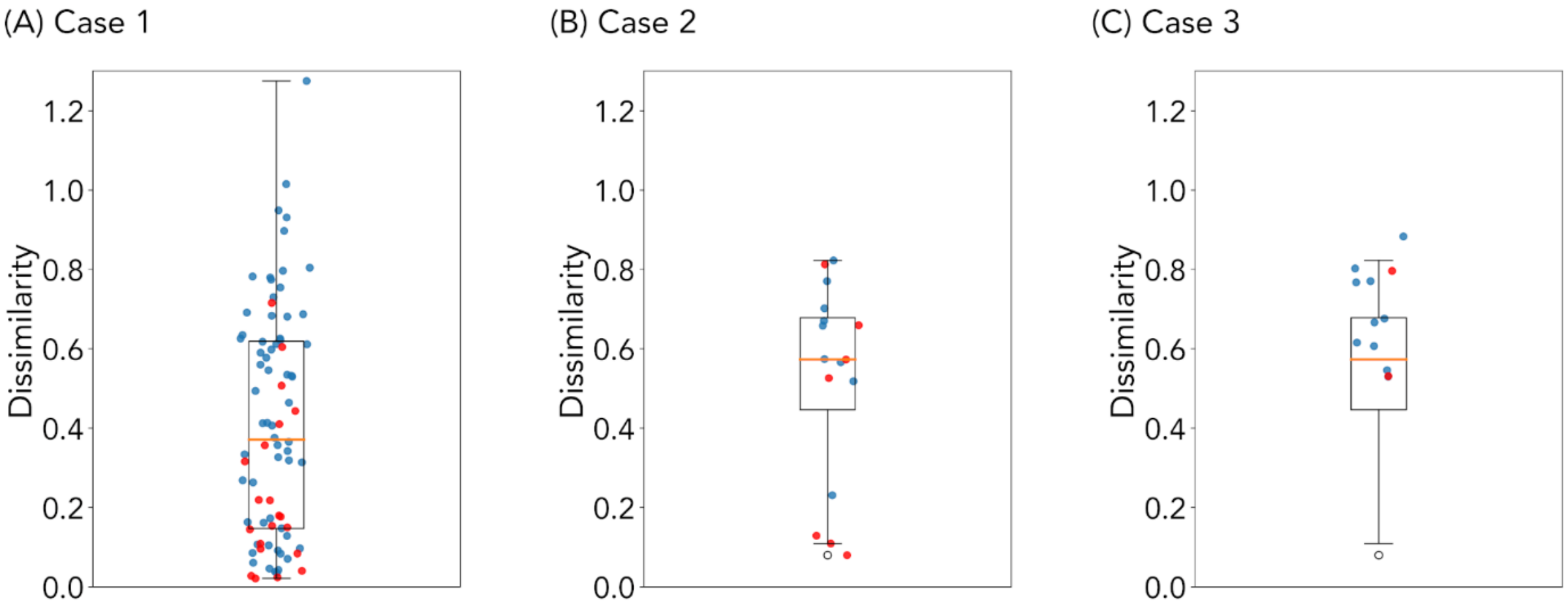

Figure S6. Dissimilarity Values of FTCP-Designed Crystals

(A) Case 1, (B) Case 2, and (C) Case 3. Red dots represent target-satisfying designed crystals.

We use dissimilarity values to assess the structural uniqueness of designed crystals. The dissimilarity value is the vector distance between two structures based on local coordination information from all sites in the two structures[28]. A zero indicates two identical crystals, while large values (>1) indicating huge dissimilarity. Figure S6 shows the minimum dissimilarity value compared to every crystal in the respective training sets for (A) Case 1, (B) Case 2, and (C) Case 3. For Case 1, the median dissimilarity value is 0.37, with 11 values (out of 84 valid designed crystals) above 0.75, which is the cutoff dissimilarity value used by Materials Project. For Case 2, the median dissimilarity value is 0.57, with three values above 0.75 (out of 16 valid designed crystals). For Case 3, the median dissimilarity value is 0.67, with five values above 0.75 (out of 12 valid designed crystals).

# S5. Invariance Study of the FTCP Representation

Table S8. Performance Degradation of the FTCP Representation Used as Property Prediction/Mapping due to Translation, Rotation, Permutation, and Different Supercells

The values are mean absolute error (MAE) | performance drop in percentage. The results are calculated with an updated version of Materials Project accessed on 14 Sep 2021. (Materials Project has updated since our design cases, which accessed Materials Project on 22 Jun 2020.) All results are mean values from five-fold cross validation. Dataset used is the same as the one in Case 1, described in Table S4.

| | Formation Energy $E_f$ (eV/atom) | Bandgap $E_g$ (eV) |
|---|---|---|
| **Data** | 0.051 | 0% (base) | 0.204 | 0% (base) |
| **Data with translation** | 0.055 | 9.1% | 0.223 | 9.3% |
| **Data with rotation** | 0.053 | 3.4% | 0.216 | 5.9% |
| **Data with permuted sites** | 0.242 | 376.3% | 0.418 | 104.8% |
| **Data with different supercells** | 0.074 | 44.8% | 0.262 | 28.5% |

The FTCP representation satisfies no invariances. To provide quantification to the performance drop due to different invariance operations, such as translation, rotation, permutation (of the order of sites), and different supercells, we offer Table S8 to evaluate the performance drop of the FTCP representation used for property prediction/mapping (although the FTCP representation is mainly developed to do inverse design). Translation is to apply translation to the site fractional coordinates. Rotation, in our case, is to apply permutation to columns of the lattice matrix, and the site coordinate matrix, *i.e.*, exchanging a, b, or c. Permutation is to apply the permutation to rows of the site coordinate matrix, and the site occupancy matrix, *i.e.*, the order of site inputted. Different supercells are to construct a new supercell such that the new lattice vectors are a linear combination of the original ones, while preserving a maximum number of sites to be 20. We observe that the FTCP representation respond most prominently to permutation of sites, resulting in the largest performance drop.